\documentclass[preprint,prd,aps,showpacs,nofootinbib]{revtex4}
\usepackage{inputenc}
\usepackage{epsfig,float}
\usepackage{amsfonts}
\usepackage[hypertex]{hyperref}
\usepackage{longtable, array}
\usepackage{amsmath}
\input{epsf}
\newcommand{\be}{\begin{eqnarray}}
\newcommand{\ee}{\end{eqnarray}}
\newcommand{\ba}{\begin{array}}
\newcommand{\ea}{\end{array}}
\newcommand{\bi}{\begin{itemize}}
\newcommand{\ei}{\end{itemize}}

\restylefloat{figure}
\restylefloat{table}

\begin{document}

\title{QCD description of backward vector meson hard electroproduction}

\author{B.~Pire$^1$,  K.~Semenov-Tian-Shansky$^{1,2}$, L.~Szymanowski$^{3}$ }
\affiliation{
$^1$ CPhT, \'{E}cole Polytechnique, CNRS,  91128 Palaiseau, France  \\
$^2$ IFPA, d\'{e}partement AGO,  Universit\'{e} de  Li\`{e}ge, 4000 Li\`{e}ge,  Belgium \\
$^3$ National Centre for Nuclear Research (NCBJ), Warsaw, Poland \\
}

\preprint{CPHT-RR004.0315}

\begin{abstract}
 We consider backward vector meson exclusive electroproduction off nucleons in the framework of collinear QCD factorization.
 Nucleon to vector meson transition distribution amplitudes arise as building blocks for the corresponding
 factorized amplitudes. In the near-backward kinematics, the suggested factorization mechanism results
 in the dominance of the transverse cross section  of vector meson production
 ($\sigma_T \gg \sigma_L$)
 and in the characteristic $1/Q^8$-scaling behavior of the cross section.
 We evaluate nucleon to vector meson TDAs in the cross-channel nucleon exchange model
 and present estimates of the differential cross section for backward
 $\rho^0$,
 $\omega$
 and
 $\phi$
 meson production off protons. The resulting cross sections are shown to be measurable in the forthcoming JLab@$12$ GeV experiments.
\end{abstract}

\maketitle
\thispagestyle{empty}
\renewcommand{\thesection}{\arabic{section}}
\renewcommand{\thesubsection}{\arabic{subsection}}

\section{Introduction}
\label{Sec_Intro}

The factorization of exclusive amplitudes into a short distance dominated part - the coefficient function - calculable in a
perturbative way on the one hand, and universal  hadronic matrix elements of non-local operators on the light-cone on the other hand,
is a key feature of quantum chromodynamics (QCD).
This allows one to extract information on the hadronic structure from measurements of specific exclusive processes in specific
kinematics. The textbook examples of such  factorization
\cite{Collins:1996fb,Radyushkin:1997ki}
are the nearly forward deeply virtual Compton scattering (DVCS)
and meson hard electroproduction, where generalized parton (quark and gluon)
distributions (GPDs) are the relevant hadronic matrix elements.
The extension of this strategy to other processes, such
as backward meson hard electroproduction
and the cross conjugated nucleon-antinucleon annihilation into a lepton pair in association with a light meson,
has been advocated in
\cite{Frankfurt:1999fp,Frankfurt:2002kz,Pire:2005ax,Lansberg:2007ec,Lansberg:2007se}
 -
although the corresponding factorization theorems are not yet rigorously proven.
For this latter class of hard  exclusive process, new hadronic matrix elements of three quark
operators on the light cone, the baryon-to-meson transition distribution amplitudes (TDAs), appear.
Baryon-to-meson TDAs  share common features both with baryon
DAs (that are defined as the baryon-to-vacuum matrix elements of the same three quark light-cone operators)
and with GPDs, since the matrix element in question depends on the longitudinal momentum transfer between a baryon and a meson
characterized by the skewness variable
$\xi$.
Also, similarly to GPDs
\cite{Impact1,Impact2,Impact3},
switching to the impact parameter space through the Fourier transform in $\Delta_T$ brings
a novel transverse picture of the nucleon. It encodes new valuable complementary information on the
hadronic $3$-dimensional structure, whose detailed physical meaning still awaits its clarification.

A collinear factorized description for backward reactions requires
the presence of a large scale
$Q$,
to ensure the  perturbative expansion of the hard subprocess in the QCD coupling constant
$\alpha_s(\mu)$
at the factorization scale
$\mu =O(Q)$.
The large scale $Q$ can be taken either as the space-like virtuality of the electromagnetic
probe in the case of electroproduction processes
\cite{Lansberg:2007ec},
or, respectively, time-like virtuality of a photon
(or mass of heavy quarkonium) for the case of cross conjugated processes with
lepton pair emission (or heavy quarkonium production) in association with a light meson
in antinucleon nucleon annihilation
\cite{Lansberg:2007se,Pire:2013jva,Pire:2013tpa}.

Our previous studies
\cite{Lansberg:2007ec,Lansberg:2007se,Pire:2011xv,Lansberg:2011aa,Lansberg:2012ha,Pire:2013jva}
were almost exclusively restricted to the case of
$\pi N$
TDAs.
Thanks to the chiral properties of QCD, $\pi N$  TDAs possess a well-understood  soft pion limit,
which can be related to the
$\xi \to 1$
limit of the TDAs. This easily allows us to work out the physical normalization of $\pi N$ TDAs
and is helpful for practical model building.

The special  $\pi N$  TDA case however does not exhaust all interesting possibilities, and
the vector meson sector should be experimentally accessible as well as the pseudoscalar meson sector
\cite{Kub,Ma:2014pka,Ma2,Singh:2014pfv}.
In this paper we consider nucleon-to-vector meson ($VN$) TDAs and address the possibility of accessing
them experimentally through backward hard electroproduction  reactions.
The spin-$1$ nature of the produced mesons gives rise to new structures for TDAs.
This enables to define a set of the leading twist-$3$ $VN$ TDAs,
which in principle may be accessed separately through a rich variety of polarization dependent observables.
In the present study we enlarge the scope of nucleon to meson distributions to the cases of
$\rho^0$, $\omega$ and $\phi$ mesons.
These three cases share the same
$J^{PC}$
quantum numbers, but each of them addresses a specific question.
The
$\phi$-meson
case deals with the issue of the strangeness content of the nucleon, which has been the subject of many experimental and theoretical studies
(see {\it e.g.}
\cite{Shanahan:2014tja} and Refs. therein),
while the combined analysis  of $\rho$ and $\omega$ production allows one to disentangle
the isotopic structure of $VN$ TDAs in the non-strange sector.

The paper is organized as follows:
in Section~\ref{Sec_Kinematics}
we describe  the kinematics of backward meson electroproduction.
In Section~\ref{Sec_Par_TDAs}
we propose a parametrization of nucleon-to-vector-meson TDAs.
We calculate the hard amplitude in
Section~\ref{Sec_ampl_calc}.
In Section~\ref{Sec_CS} we
compute the unpolarized cross-section for backward hard meson electroproduction off nucleons
and present estimates for the cross section of the backward
$\omega(782)$,
$\phi(1020)$
and
$\rho^0(770)$
production
within the cross-channel nucleon exchange model of the
$VN$
TDAs.
Section~\ref{Sec_Concl}
brings our conclusions.
Appendix~\ref{App_Isospin}
explores the isospin constraints  and the permutation properties of the
$VN$
TDAs;
appendix~\ref{App_NPole}
derives the cross channel nucleon exchange model for the
$VN$
TDAs.

\section{Kinematics of backward vector meson hard electroproduction}
\label{Sec_Kinematics}

\begin{figure}[h]
 \centering
 \epsfig{figure= 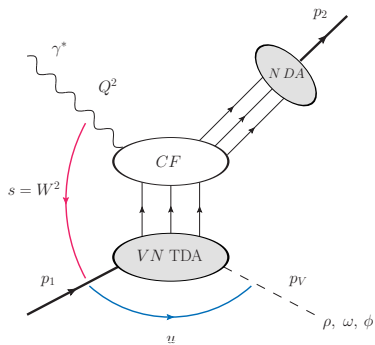 , height=4.5cm}
  \caption{ Collinear factorization  of $\gamma^* N \to N V$  in the  near-backward  kinematics regime (large $Q^2$, $s$; fixed $x_{B}$; $|u| \sim 0$);
  $VN$ TDA stands for the transition
     distribution amplitude from a nucleon to a vector meson; $N$ DA stands for the nucleon distribution amplitude;
      $CF$  denotes hard subprocess amplitudes (coefficient functions).  }
\label{Fig_TDAfact}
\end{figure}

\begin{figure}[h]
 \begin{center}
 \epsfig{figure= 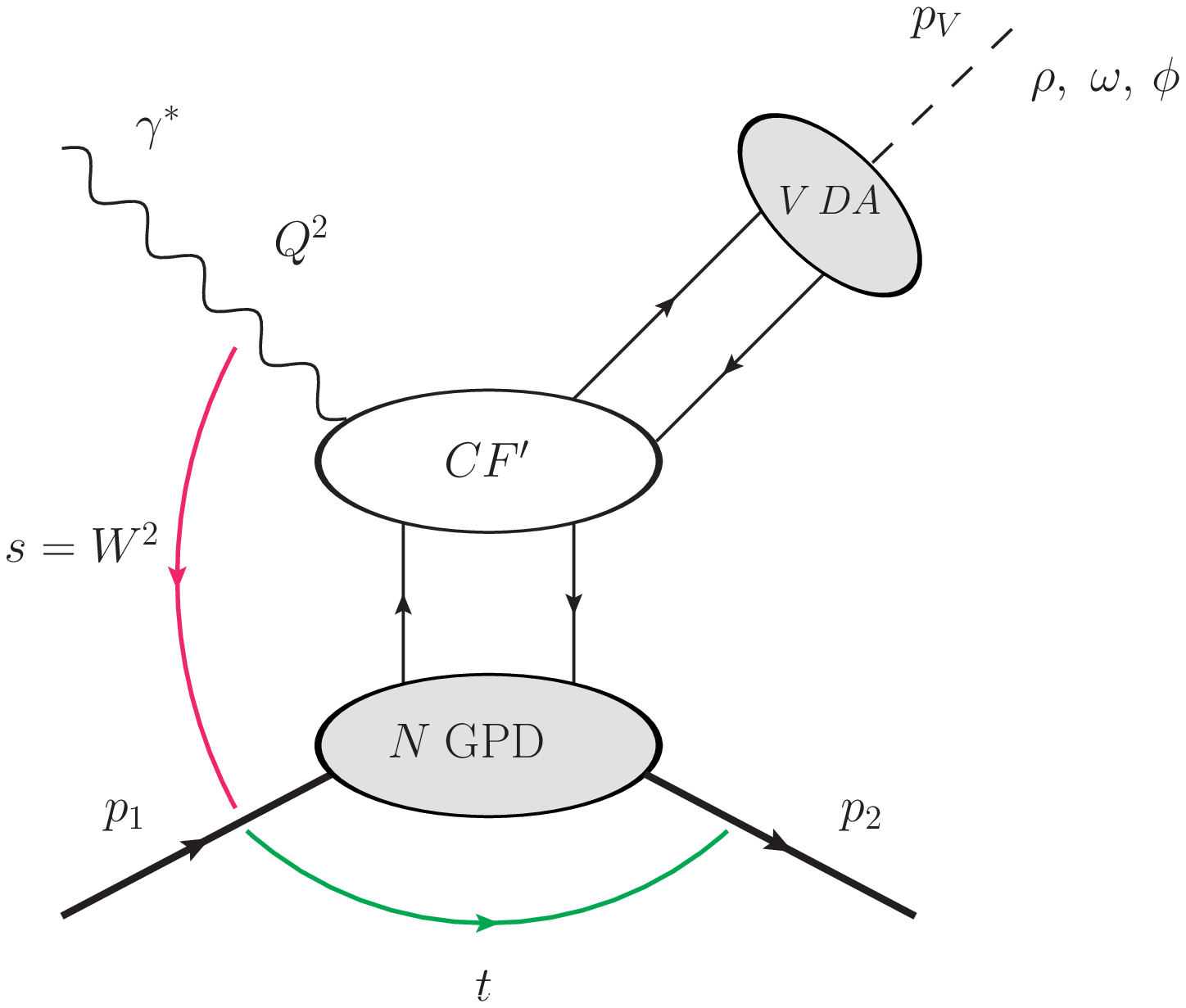 , height=4.5cm}
  \epsfig{figure= 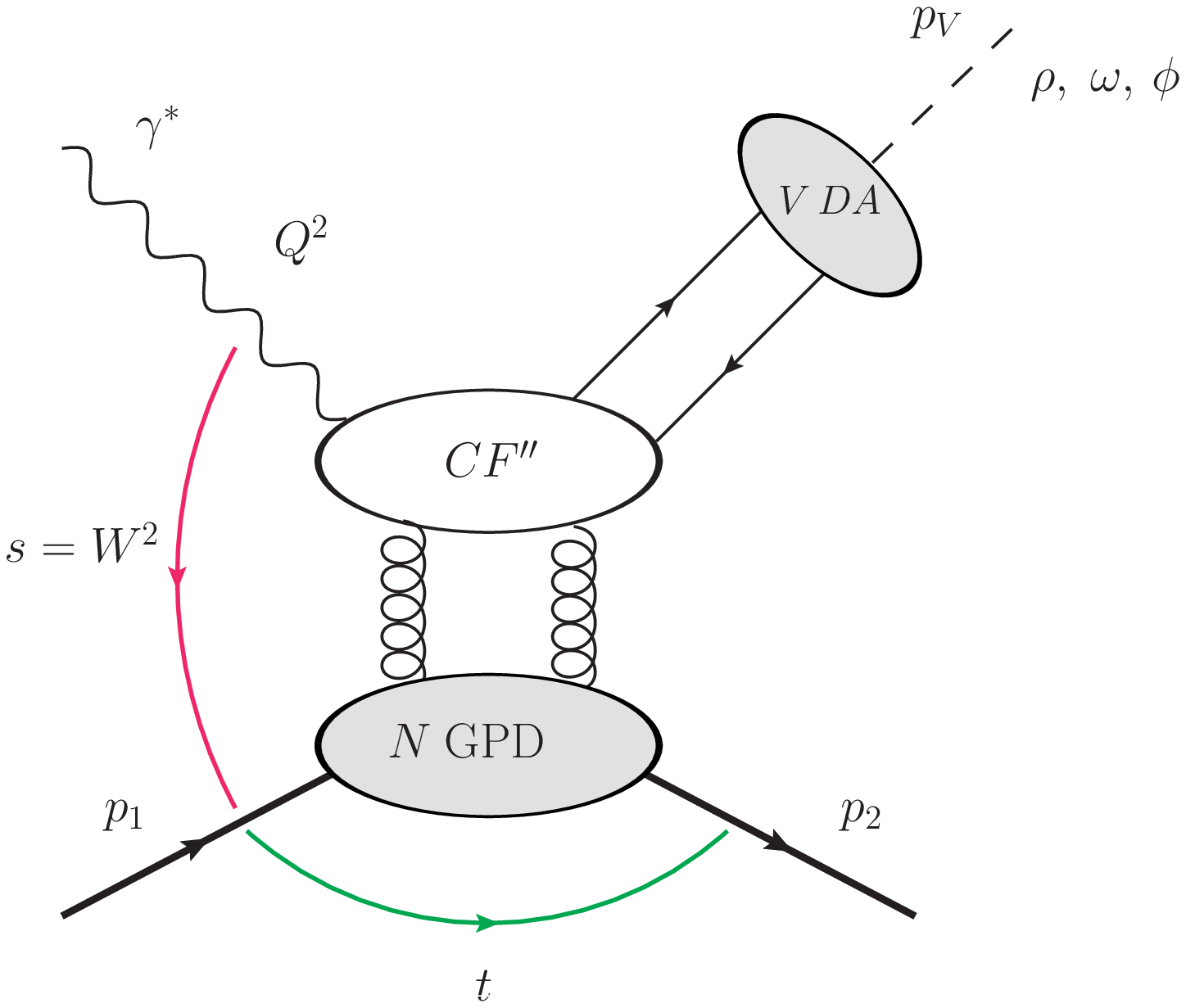 , height=4.5cm}
 \end{center}
  \caption{ Collinear factorization  of $\gamma^* N \to N V$  in the  near-forward  kinematics regime (large $Q^2$, $s$; fixed $x_{B}$; $|t| \sim 0$);
  $N$ GPD stand for, respectively, quark and gluon nucleon GPDs; $V$ DA stands for the vector meson distribution amplitude;
      $CF'$ and $CF''$ denote the corresponding hard subprocess amplitudes.  }
\label{Fig_GPDfact}
\end{figure}

We consider the exclusive electroproduction of vector mesons off nucleons
\be
e(k)+N(p_1,s_1) \to
\left(\gamma^*(q, \lambda_\gamma)+N(p_1,s_1) \right) + e(k') \to e(k')+N(p_2,s_2)+ V(p_V, \lambda_V),
\label{hard_omega_production}
\ee
within the generalized Bjorken limit, in which
$Q^2=-q^2$ and $s$ are large%
\footnote{Throughout this paper we employ the usual Mandelstam variables for the
hard subprocess of the reaction
(\ref{hard_omega_production}):
$s=(p_1+q)^2 \equiv W^2$; $t=(p_2-p_1)^2$; $u=(p_V-p_1)^2 \equiv \Delta^2$.
};
the Bjorken variable $x_{B} \equiv \frac{Q^2}{2 p_1 \cdot q}$ is fixed and
the $u$-channel momentum transfer squared is small compared to $Q^2$ and $s$:
$|u| \equiv |\Delta^2| \ll Q^2,\,s$.
Within such kinematics, the amplitude of the hard subprocess of the reaction
(\ref{hard_omega_production})
is supposed to admit a collinear factorized description in terms of
nucleon-to-vector-meson TDAs and nucleon DAs, as it is shown on Fig.~\ref{Fig_TDAfact}.
The small $u$ corresponds to the vector meson produced in the near-backward direction in the
$\gamma^* N$ center-of-mass system (CMS).
Therefore in what follows, we refer the kinematical regime in question as the near-backward kinematics.
We would like to emphasize that this kinematical regime is complementary to the more familiar
generalized Bjorken limit
($Q^2$ and $s$ - large; $x_{B}$ - fixed; $|t| \ll Q^2,\,s$),
known as the near-forward kinematics.
In this latter kinematical regime the conventional collinear factorization theorem
\cite{Collins:1996fb,Radyushkin:1997ki}
leading to the description of
(\ref{hard_omega_production})
in terms of GPDs and vector meson DAs is established
 (see Fig.~\ref{Fig_GPDfact}).

We choose the $z$-axis along the colliding virtual-photon-nucleon
and introduce the light-cone vectors
$p$
and
$n$
($2 p \cdot n=1$).
Keeping the first-order corrections in the masses and
$\Delta_T^2$,
we establish the following Sudakov decomposition for the momenta of the reaction
(\ref{hard_omega_production})
in the near-backward kinematical regime
({\it cf.} \cite{Lansberg:2007ec}):
\be
&&
\nonumber
p_1 = (1+\xi) p + \frac{M^2}{1+\xi}n \,; \nonumber \\ &&
\nonumber
p_2 \simeq -2\xi \frac{(\Delta_T^2-M^2)}{Q^2} p + \left[ \frac{Q^2}{2\xi \Big(1+   \frac{(\Delta_T^2-M^2)}{Q^2}\Big)} - \frac{m^2_V-\Delta_T^2}{1-\xi}+ \frac{M^2}{1+\xi}\right]n-\Delta_T;
\nonumber \\ &&
q \simeq- 2 \xi \Big(1+ \frac{(\Delta_T^2-M^2)}{Q^2}\Big)  p + \frac{Q^2}{2\xi\Big(1+   \frac{(\Delta_T^2-M^2)}{Q^2}\Big)} n\,;  \nonumber \\ &&
\nonumber
p_V = (1-\xi) p +\frac{m^2_V-\Delta_T^2}{1-\xi}n+ \Delta_T \, ;
\\ &&
\Delta = - 2 \xi p +\Big[\frac{m_V^2-\Delta_T^2}{1-\xi}- \frac{M^2}{1+\xi}\Big]n
+ \Delta_T \,.
\label{Sudakov_dec_omega}
\ee
Here
$M$
and
$m_V$
denote respectively the nucleon and the vector meson masses and
$\xi $
stands for the $u$-channel skewness parameter introduced with respect to the $u$-channel momentum transfer
\be
\xi = -\frac{(p_V-p_1) \cdot n}{(p_V+p_1) \cdot n},
\ee
and
\be
\Delta_T^2= \frac{1-\xi}{1+\xi}\left( \Delta^2-2 \xi \left[ \frac{M^2}{1+\xi} - \frac{m_V^2}{1-\xi} \right] \right)
\label{Delta_t2}
\ee
is the $u$-channel transverse momentum transfer squared.
We introduce $u_{0}$ corresponding to $\Delta_T^2=0$:
\be
u_{0}=-\frac{2 \xi  \left(m_V^2 (1+\xi)-M^2 (1-\xi)\right)}{1-\xi ^2}.
\ee
It is the maximal possible value of $u$ for given $\xi$. For $\Delta_T^2=0$ ($u=u_0$) the vector meson is produced exactly
in the backward direction in the $\gamma^* N$ CMS ($\theta_V^*=\pi$).

In the initial nucleon rest frame (corresponding to the JLab laboratory frame, (LAB)) the
light-cone vectors $p$ and $n$ read
\be
p|_{\rm LAB}=\frac{M}{2(1+\xi)} \{1,0,0,-1\}; \ \ \ n|_{\rm LAB}=\frac{1+\xi}{2M} \{1,0,0,1\}.
\ee
With the help of the appropriate boost
we establish the expressions for the light-cone vectors
in the
$\gamma^* N$
CMS:
\be
p|_{\gamma^* N \,  {\rm CMS}}=  \{\alpha,0,0,-\alpha\}; \ \ \ p|_{\gamma^* N \, {\rm CMS}}=  \{\beta,0,0,\beta\},
\ee
with
\be
&&
\alpha= \frac{M^2+Q^2+W^2+\Lambda \left(W^2,-Q^2,M^2\right)}{4
   (1+\xi) W};
\nonumber \\ &&
\beta=\frac{(1+\xi) \left(M^2+Q^2+W^2-\Lambda
   \left(W^2,-Q^2,M^2\right)\right)}{4 M^2
   W},
\ee
where $\Lambda$
is the usual Mandelstam function
\be
\Lambda(x,y,z)= \sqrt{x^2+y^2+z^2-2xy-2xz-2yz} \, .
\label{Def_lambda}
\ee

The $V$-meson  scattering angle in the $\gamma^* N$ CMS for the $u$-channel factorization regime then can be expressed as:
\be
\cos \theta_{V}^*=\frac{-(1-\xi )\alpha +\frac{m^2_V-{\Delta_T}^2}{1-\xi} \beta}
{\sqrt{(-(1-\xi)\alpha+\frac{m^2_V-{\Delta_T}^2}{1-\xi} \beta)^2-{\Delta_T}^2}} \, \,.
\label{CosThetaV_CMS}
\ee
One may check that for ${\Delta_T}^2=0$ indeed $\cos \theta_V^*=-1$, which means backward
scattering.

For skewness variable $\xi$ we employ the approximate expression neglecting order of mass and $\Delta_T^2$  corrections
\be
\xi \simeq \frac{x_{B}}{2-x_{B}} \simeq \frac{Q^2}{Q^2+2W^2}.
\ee

From the transversality of the polarization vector of the vector meson
\be
{\cal E}^*(p_V, \lambda_V) \cdot p_V=0
\ee
we establish the following condition for the ``$-$''-light-cone component of the polarization vector of the vector meson:
\be
{\cal E}^*(p_V, \lambda_V) \cdot p= - \frac{m_V^2-\Delta_T^2}{(1-\xi)^2} ({\cal E}^*(p_V, \lambda_V) \cdot n)- \frac{1}{1-\xi} ({\cal E}^*(p_V, \lambda_V) \cdot \Delta_T).
\label{transvers_E}
\ee

\section{Parametrization of  nucleon-to-vector meson transition distribution amplitudes}
\label{Sec_Par_TDAs}

Nucleon-vector-meson TDAs are formally defined as the matrix elements of the light-cone
three quark operator between a nucleon and a vector meson states.
For simplicity we leave the discussion of isotopic spin properties of $VN$ TDAs to the
Appendix~\ref{App_Isospin}
and consider the transition matrix element of the
$uud$
light-cone operator%
\footnote{We adopt the light-cone gauge
$A^+=0$
and therefore the gauge link is not shown explicitly in the operator definition.}
\be
\hat{O}_{\rho \tau \chi}^{uud}(\lambda_1n,\lambda_2n,\lambda_3n) \equiv
\varepsilon_{c_1 c_2 c_3} u_\rho^{c_1 }(\lambda_1n) u_\tau^{c_2  }(\lambda_2n) d_\chi^{c_3}(\lambda_3n)
\ee
between the proton
$|N_p \rangle$
state and a
$I_3=0$,
$J^{PC}=1^{--}$
vector meson
state (these could be {\it e.g.}
$\langle \omega(782)|$,
$\langle \rho^0(770)|$ or
$\langle \phi(1020)| $ meson states).

The parametrization for the leading twist $V N$ TDA involves $24$ Dirac structures.
Indeed, each of the three quarks and the nucleon have $2$ helicity states, while the vector meson
has $3$. This leads to $3 \cdot 2^4=48$ helicity amplitudes. However, parity relates helicity amplitudes
with all opposite helicities reducing the overall number of independent helicity amplitudes
by the factor of $2$.
The procedure for building the corresponding leading twist Dirac structures was described in Ref.~\cite{Lansberg:2006uh}.
The revised version%
\footnote{The original parametrization of Ref.~\cite{Lansberg:2006uh} erroneously lacked
$\gamma_5$
factors for the Dirac structures. {\it C.f.} eq.~(14) of Ref.~\cite{Lansberg:2006uh}  and eqs.~(\ref{Dirac_v_NV})-(\ref{Dirac_t_NV}) of the present paper. }
 of the leading twist-$3$ $VN$ TDA parametrization reads
\be
&&
4 {\cal F} \langle V(p_V, \lambda_V) | \hat{O}_{\rho \tau \chi}^{uud}(\lambda_1n,\lambda_2n,\lambda_3n)|N^p(p_1,s_1) \rangle
\nonumber \\ &&
=
\delta(x_1+x_2+x_3-2\xi)  \times M \Big[
\sum_{\Upsilon= 1 \varepsilon, 1T, 1n, \atop 2 \varepsilon, 2T, 2n  } (v^{V N}_\Upsilon)_{\rho \tau, \, \chi} V_{\Upsilon}^{VN}(x_1,x_2,x_3, \xi, \Delta^2)
\nonumber \\ &&
+\sum_{\Upsilon= 1 \varepsilon, 1T, 1n, \atop 2 \varepsilon, 2T, 2n  } (a^{V N}_\Upsilon)_{\rho \tau, \, \chi} A_{\Upsilon}^{VN}(x_1,x_2,x_3, \xi, \Delta^2)
+
\sum_{\Upsilon= 1 \varepsilon, 1T, 1n,   \varepsilon, 2T, 2n, \atop 3 \varepsilon, 3T, 3n,  4 \varepsilon, 4T, 4n } (t^{V N}_\Upsilon)_{\rho \tau, \, \chi} T_{\Upsilon}^{VN}(x_1,x_2,x_3, \xi, \Delta^2)
\Big],
\nonumber \\ &&
\label{VN_TDAs_param}
\ee
where $\cal F$ stands for the conventional Fourier transform
\be
{\cal F} \equiv {\cal F}(x_1,x_2,x_3)(\ldots)= (P \cdot n)^3 \int \left[ \prod_{j=1}^3 \frac{d \lambda_j}{2 \pi} \right] e^{i \sum_{k=1}^3 x_k \lambda_k (P \cdot n)} \,,
\label{Fourier_TDA}
\ee
and the leading twist Dirac structures are defined as
\be
&&
(v_{1 {\cal E}}^{V N})_{\rho \tau, \, \chi}= (\hat{p}C)_{\rho \tau} \big(\gamma^5 \hat{{\cal E}}^*(p_V, \lambda_V)U^+(p_1,s_1) \big)_\chi; \nonumber \\ &&
(v_{1 T}^{V N})_{\rho \tau, \, \chi}=
M^{-1}
({{\cal E}}^*(p_V, \lambda_V)
\cdot \Delta_T)
(\hat{p}C)_{\rho \tau} \big(\gamma^5 U^+(p_1,s_1) \big)_\chi; \nonumber \\ &&
(v_{1 n}^{V N})_{\rho \tau, \, \chi}= M  ({{\cal E}^*}(p_V, \lambda_V)
\cdot n) (\hat{p}C)_{\rho \tau} \big(\gamma^5 U^+(p_1,s_1) \big)_\chi; \nonumber \\ &&
(v_{2 \varepsilon}^{V N})_{\rho \tau, \, \chi}=  M^{-1}   (\hat{p}C)_{\rho \tau} \big(\gamma^5  \sigma^{\Delta_T {\cal E}^* } U^+(p_1,s_1) \big)_\chi;
\nonumber \\ &&
(v_{2 T}^{V N})_{\rho \tau, \, \chi}=  M^{-2}
({{\cal E}}^*(p_V, \lambda_V) \cdot \Delta_T)
(\hat{p}C)_{\rho \tau}  \big(\gamma^5 \hat{\Delta}_T U^+(p_1,s_1) \big)_\chi; \nonumber \\ &&
(v_{2 n}^{V N})_{\rho \tau, \, \chi}=
({{\cal E}}^*(p_V, \lambda_V) \cdot n)
(\hat{p}C)_{\rho \tau}  \big(\gamma^5 \hat{\Delta}_T U^+(p_1,s_1) \big)_\chi;
\label{Dirac_v_NV}
\ee
\be
&&
(a_{1 \varepsilon}^{V N})_{\rho \tau, \, \chi}= (\hat{p} \gamma^5 C)_{\rho \tau} \big(  \hat{{\cal E}}^*(p_V, \lambda_V) U^+(p_1,s_1) \big)_\chi; \nonumber \\ &&
(a_{1 T}^{V N})_{\rho \tau, \, \chi}=
M^{-1}
({{\cal E}}^*(p_V, \lambda_V)
\cdot \Delta_T)
(\hat{p} \gamma^5 C)_{\rho \tau} \big(  U^+(p_1,s_1) \big)_\chi; \nonumber \\ &&
(a_{1 n}^{V N})_{\rho \tau, \, \chi}= M  ({{\cal E}}^*(p_V, \lambda_V)
\cdot n) (\hat{p} \gamma^5 C)_{\rho \tau} \big(  U^+(p_1,s_1) \big)_\chi; \nonumber \\ &&
(a_{2 \varepsilon}^{V N})_{\rho \tau, \, \chi}=  M^{-1}   (\hat{p} \gamma^5C)_{\rho \tau} \big(   \sigma^{\Delta_T {\cal E}^* } U^+(p_1,s_1) \big)_\chi;
\nonumber \\ &&
(a_{2 T}^{V N})_{\rho \tau, \, \chi}=  M^{-2}
({{\cal E}}^*(p_V, \lambda_V) \cdot \Delta_T)
(   \hat{p}\gamma^5 C)_{\rho \tau}  \big(\hat{\Delta}_T U^+(p_1,s_1) \big)_\chi; \nonumber \\ &&
(a_{2 n}^{V N})_{\rho \tau, \, \chi}=
({{\cal E}}^*(p_V, \lambda_V) \cdot n)
(   \hat{p} \gamma^5  C)_{\rho \tau}  \big(\hat{\Delta}_T U^+(p_1,s_1) \big)_\chi;
\label{Dirac_a_NV}
\ee
\be
&&
(t_{1 \varepsilon}^{V N})_{\rho \tau, \, \chi}=(\sigma_{p \lambda} C)_{\rho \tau} (\gamma_5 \sigma^{\lambda {\cal E}^*}  U^+(p_1,s_1))_\chi;
\nonumber \\ &&
(t_{1 T}^{V N})_{\rho \tau, \, \chi}= M^{-1}  ({\cal E}^*(p_V, \lambda_V) \cdot \Delta_T) (\sigma_{p \lambda} C)_{\rho \tau} (\gamma_5 \gamma^\lambda  U^+(p_1,s_1))_\chi;
\nonumber \\ &&
(t_{1 n}^{V N})_{\rho \tau, \, \chi}= M   ({\cal E}^*(p_V, \lambda_V) \cdot n) (\sigma_{p \lambda} C)_{\rho \tau} (\gamma_5 \gamma^\lambda  U^+(p_1,s_1))_\chi;
\nonumber \\ &&
(t_{2 \varepsilon}^{V N})_{\rho \tau, \, \chi}=(\sigma_{p {\cal E}^*} C)_{\rho \tau} (\gamma_5   U^+(p_1,s_1))_\chi;
\nonumber \\ &&
(t_{2 T}^{V N})_{\rho \tau, \, \chi}= M^{-2}  ({\cal E}^*(p_V, \lambda_V) \cdot \Delta_T) (\sigma_{p \lambda} C)_{\rho \tau} (\gamma_5 \sigma^{\lambda \Delta_T}  U^+(p_1,s_1))_\chi;
\nonumber \\ &&
(t_{2 n}^{V N})_{\rho \tau, \, \chi}= ({\cal E}^*(p_V, \lambda_V) \cdot n) (\sigma_{p \lambda} C)_{\rho \tau} (\gamma_5 \sigma^{\lambda \Delta_T}  U^+(p_1,s_1))_\chi;
\nonumber \\ &&
(t_{3 \varepsilon}^{V N})_{\rho \tau, \, \chi}=M^{-1} (\sigma_{p \Delta_T} C)_{\rho \tau} (\gamma_5 \hat{{\cal E}}^*(p_V, \lambda_V) U^+(p_1,s_1))_\chi;
\nonumber \\ &&
(t_{3 T}^{V N})_{\rho \tau, \, \chi}= M^{-2}  ({\cal E}^*(p_V, \lambda_V) \cdot \Delta_T) (\sigma_{p \Delta_T} C)_{\rho \tau} (\gamma_5  U^+(p_1,s_1))_\chi;
\nonumber \\ &&
(t_{3 n}^{V N})_{\rho \tau, \, \chi}=   ({\cal E}^*(p_V, \lambda_V) \cdot n) (\sigma_{p \Delta_T} C)_{\rho \tau} (\gamma_5  U^+(p_1,s_1))_\chi;
\nonumber \\ &&
(t_{4 \varepsilon}^{V N})_{\rho \tau, \, \chi}= M^{-1} (\sigma_{p {\cal E}^*} C)_{\rho \tau} (\gamma_5 \hat{\Delta}_T U^+)_\chi;
\nonumber \\ &&
(t_{4 T}^{V N})_{\rho \tau, \, \chi}=M^{-3} ({\cal E}^*(p_V, \lambda_V) \cdot \Delta_T) (\sigma_{p \Delta_T} C)_{\rho \tau} (\gamma_5 \hat{\Delta}_T U^+(p_1,s_1))_\chi;
\nonumber \\ &&
(t_{4 n}^{V N})_{\rho \tau, \, \chi}=M^{-1}  ({\cal E}^*(p_V, \lambda_V) \cdot n)   (\sigma_{p \Delta_T} C)_{\rho \tau} (\gamma_5 \hat{\Delta}_T U^+(p_1,s_1))_\chi.
\label{Dirac_t_NV}
\ee
Throughout this paper we employ Dirac's ``hat'' notation:
$\hat{a} \equiv \gamma_{\mu} a^{\mu}$
and adopt the usual conventions:
$\sigma^{\mu \nu}= \frac{1}{2} [\gamma^\mu, \, \gamma^\nu]$;
$\sigma^{v \nu} \equiv v_\mu \sigma^{\mu \nu}$,
where $v_\mu$ is an arbitrary $4$-vector.
The large and small components of the nucleon Dirac spinor $U(p_1)$
are introduced as
$U^+(p_1,s_1)=  \hat{p} \hat{n}  U(p_1,s_1)$
and
$U^-(p_1,s_1)=  \hat{n} \hat{p}  U(p_1,s_1)$.

Each of the $24$ $VN$ TDAs defined in
(\ref{VN_TDAs_param})
are functions of three longitudinal momentum fractions $x_1$, $x_2$, $x_3$, skewness parameter $\xi$, $u$-channel
momentum transfer squared $\Delta^2$ and of factorization scale $\mu^2$.
TDAs
$V_{\Upsilon}^{VN}(x_1,x_2,x_3, \xi, \Delta^2)$
and
$T_{\Upsilon}^{VN}(x_1,x_2,x_3, \xi, \Delta^2)$
are defined symmetric under the interchange $x_1 \leftrightarrow x_2$,
while
$A_{\Upsilon}^{VN}(x_1,x_2,x_3, \xi, \Delta^2)$
are antisymmetric under the interchange $x_1 \leftrightarrow x_2$.

Note that with the use of the parametrization
(\ref{VN_TDAs_param})
the corresponding TDAs do not satisfy the polynomiality property in its simple form.
Indeed, as explained in Ref.~\cite{Lansberg:2011aa}, since the light-cone kinematics implies
the choice of a preferred $z$-direction, the parametrization
(\ref{VN_TDAs_param})
involves non-covariant kinematical quantities (such as $\Delta_T$ and light-cone vectors $p$ and $n$).
This results in the presence of kinematical singularities for the corresponding invariant amplitudes (TDAs).
In principle, one can define the alternative set of the Dirac structures for
$VN$
involving only fully covariant kinematical quantities such as four-vectors
$P=\frac{1}{2} (p_1+p_V)$
and
$\Delta$. Thus, for the price of the controllable admixture of higher twist contributions
the corresponding set of
$VN$
TDAs turns to be free of kinematical singularities and satisfies the polynomiality condition in its simple form.
Therefore, for this set of TDAs one can introduce the spectral representation
\cite{Pire:2010if}
in terms of quadruple distributions.
The relation between the free-of-kinematical-singularities-set of $VN$ TDAs and those introduced in eq.~(\ref{VN_TDAs_param}) is given by the
set of relations similar to eq.~(C11) of Ref.~\cite{Lansberg:2011aa}.

In the present study we, nevertheless, prefer to stay with the $VN$ TDA parametrization
(\ref{VN_TDAs_param}),
since it is well suited to keep eye on the
$\Delta_T = 0$
limit.
Namely, in the limit $\Delta_T=0$
only
$7$
TDAs out of $24$  turn to be relevant:
$V_{1 {\cal E}}^{VN}$,  $V_{1 n}^{VN}$,
$A_{1 {\cal E}}^{VN}$,  $A_{1 n}^{VN}$, $
T_{1 {\cal E}}^{VN}$, $T_{1 n}^{VN}$,
$T_{2 {\cal E}}^{VN}$.


\section{Calculation of the hard amplitude}
\label{Sec_ampl_calc}

Within the suggested factorized approach, in the leading order (both in $\alpha_s$ and $1/Q$)
the amplitude of%
\footnote{For definiteness we take $V$ to be a vector meson with $I_3=0$: $\omega$, $\rho^0$ or $\phi$.}
\be
\gamma^*(q, \lambda_\gamma) + N^p(p_1,s_1) \to N^p(p_2,s_2) + V( p_V, \lambda_{V}  )
\label{reaction_amp}
\ee
involves $6$ independent tensor structures
\be
{\cal M}_{s_1 s_2}^{\lambda_\gamma \lambda_V}= -i \frac{(4 \pi \alpha_s)^2 \sqrt{4 \pi \alpha_{em}} f_N M}{54 Q^4}
\sum_{k=1}^6
{\cal S}_{s_1 s_2}^{(k) \, \lambda_\gamma \lambda_V} {\cal I}^{(k)} (\xi, \, \Delta^2).
\label{Hel_ampl_def}
\ee
Here $\alpha_{em}= \frac{1}{137}$  is the electromagnetic fine structure constant and $f_N$
is the nucleon light-cone wave function normalization constant. Throughout this study we
employ the value from Ref.~\cite{Chernyak:1984bm}:
$
f_N=5.2 \cdot 10^{-3} \; {\rm GeV}^2
$.

There turn to be two tensor structures independent of $\Delta_T$:
\be
&&
{\cal S}_{s_1 s_2}^{(1) \, \lambda_\gamma \lambda_V}= \bar{U}(p_2,s_2) \hat{\varepsilon}(q,\lambda_\gamma) \hat{\cal E}^*(p_V, \lambda_V) U(p_1,s_1);
\nonumber \\ &&
{\cal S}_{s_1 s_2}^{(2) \, \lambda_\gamma \lambda_V}= M ({\cal E}^*(p_V, \lambda_V) \cdot n) \bar{U}(p_2,s_2) \hat{\varepsilon}(q,\lambda_\gamma)  U(p_1,s_1),
\ee
and four $\Delta_T$-dependent tensor structures:
\be
&&
{\cal S}_{s_1 s_2}^{(3) \, \lambda_\gamma \lambda_V}= \frac{1}{M} ({\cal E}^*(p_V, \lambda_V) \cdot \Delta_T) \, \bar{U}(p_2,s_2) \hat{\varepsilon}(q,\lambda_\gamma)  U(p_1,s_1);
\nonumber \\ &&
{\cal S}_{s_1 s_2}^{(4) \, \lambda_\gamma \lambda_V}=\frac{1}{M^2} ({\cal E}^*(p_V, \lambda_V) \cdot \Delta_T) \, \bar{U}(p_2,s_2) \hat{\varepsilon}(q,\lambda_\gamma) \hat{\Delta}_T U(p_1,s_1);
\nonumber \\ &&
{\cal S}_{s_1 s_2}^{(5) \, \lambda_\gamma \lambda_V}=\frac{1}{M}   \, \bar{U}(p_2,s_2) \hat{\varepsilon}(q,\lambda_\gamma)
\hat{{\cal E}}^*(p_V, \lambda_V)
\hat{\Delta}_T U(p_1,s_1);
\nonumber \\ &&
{\cal S}_{s_1 s_2}^{(6) \, \lambda_\gamma \lambda_V}=   ({\cal E}^*(p_V, \lambda_V) \cdot n) \, \bar{U}(p_2,s_2) \hat{\varepsilon}(q,\lambda_\gamma)
\hat{{\cal E}}^*(p_V, \lambda_V)
\hat{\Delta}_T U(p_1,s_1).
\ee
Here
$ \varepsilon(q, \lambda_\gamma)$
stands for the polarization vector of the incoming virtual photon and
${\cal E}^*(p_V, \lambda_V)$
is the polarization vector of the outgoing vector meson.

To the leading order in
$\alpha_s$,
within the collinear factorized description in terms of $VN$ TDAs, the amplitude of the
reaction
(\ref{reaction_amp})
can be computed from the same $21$ diagrams listed in Table I of Ref.~\cite{Lansberg:2007ec}.
We adopt our common notations for the integral convolutions
$\mathcal{I}^{(k)}$, $k=1, \, \ldots,\, 6$:
\be
&&
\mathcal{I}^{(k)} (\xi,\Delta^2) \equiv { {\int^{1+\xi}_{-1+\xi} }\! \! \! dx_1  {\int^{1+\xi}_{-1+\xi} }\! \! \! dx_2  {\int^{1+\xi}_{-1+\xi} }\! \! \!dx_3 \, \delta(x_1+x_2+x_3-2\xi)
}
\nonumber \\ &&
\times
{{\int^{1}_{0} }\! \! \! dy_1  {\int^{1}_{0} }\! \! \! dy_2  {\int^{1}_{0} }\! \! \!dy_3 \, \delta(y_1+y_2+y_3-1)}
{\Bigg(2\sum_{\alpha=1}^{7}   T_{\alpha}^{(k)} +
\sum\limits_{\alpha=8}^{14}   T_{\alpha}^{(k)} \Bigg)}.
\label{Def_IIprime_vector}
\ee
The explicit expressions for the coefficients  $T_{\alpha}^{(k)} \equiv  D_\alpha \times N_\alpha^{(k)}$ (no summation over $\alpha$ assumed)
are presented in Table~I. Here $D_\alpha$ denote the singular kernels originating from the partonic propagators and
$N_\alpha^{(k)} \equiv N_\alpha^{(k)}(x_1,x_2,\xi, \Delta^2; \, y_1, y_2,y_3)$
stand for the appropriate combinations of $VN$ TDAs and nucleon DAs $V^p$, $A^p$ and $T^p$ arising in the numerator.
The
$\alpha=1, \ldots, 21$
index refers for the diagram number and the index
$k=1, \ldots, 6$
runs for the contributions into $6$ invariant amplitudes of eq.~(\ref{Hel_ampl_def}).

\begin{longtable}{|c|p{5.1cm} |c|}
\newpage
\caption{14 of the 21 diagrams contributing to the hard-scattering amplitude with
their associated coefficient $T_\alpha^{(k)} \equiv D_\alpha \times N_\alpha^{(k)}$ (no summation over $\alpha$ assumed).    The seven first ones with $u$-quark
lines inverted are not drawn. The crosses represent the virtual-photon vertex.} \\
\hline
$\alpha$ &  \qquad \quad Diagram  & Numerators \\
        &  \qquad \qquad  $D_\alpha$  &  \\
        \nopagebreak
\hline
1 & \raisebox{-0.0cm}
{\includegraphics[height=1.5cm,clip=true]{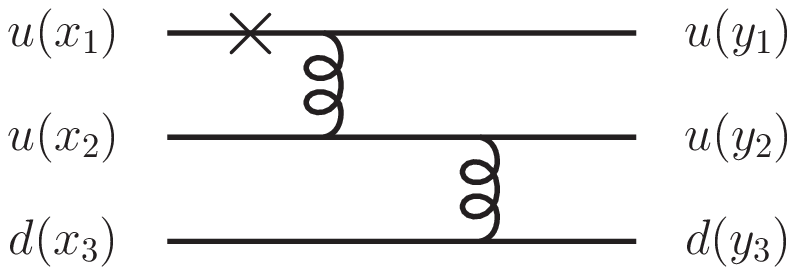}}
$\frac{Q_u (2\xi)^2}{(2\xi-x_{1}-i\epsilon)^2(x_{3}-i\epsilon)(1-y_{1})^2y_{3}}$ &
\begin{tabular}{p{0.85cm}|p{9.2cm}}
 $N_\alpha^{(1)}$ & $
-\left(V^p-A^p\right)
   \left(V_{1 {\cal E}}^{V N}+A_{1 {\cal E}}^{V N}\right)
   +
   2 T^{p} \left(T_{1 {\cal E}}^{V N}+T_{2 {\cal E}}^{V N}\right)
$ \\ \hline
 $N_\alpha^{(2)}$
 & $ -\left(V^p-A^p\right)
   \left(V_{1 n}^{V N}+A_{1 n}^{V N}\right)
   +
   4 T^{p}  \left( T_{1 n}^{V N} + \frac{\Delta_T^2 }{2 M^2} T_{4 n}^{V N} \right)
$
\\ \hline
\raisebox{-0.2cm}{  $N_\alpha^{(3)}$ } &
$
-\left(V^p-A^p\right)
   \left(V_{1 T}^{V N}+A_{1 T}^{V N}+V_{2 {\cal E}}^{V N}+A_{2 {\cal E}}^{V N}\right)
  $
\\
    &  $ +
   4 T^{p}  \left( T_{1 T}^{V N}+T_{3 {\cal E}}^{V N} + \frac{\Delta_T^2  }{2 M^2} T_{4T}^{V N} \right)$  \\ \hline
  $N_\alpha^{(4)}$ &  $
-\left(V^p-A^p\right)
   \left(V_{2 T}^{V N}+A_{2 T}^{V N} \right)
   +
   2 T^{p}  \left( T_{2 T}^{V N}+T_{3 T}^{V N}   \right)
$ \\ \hline
 $N_\alpha^{(5)}$  &  $
\left(V^p-A^p\right)
   \left(V_{2 {\cal E}}^{V N}+A_{2 {\cal E}}^{V N} \right)
   -
   2 T^{p}  \left( T_{3 {\cal E}}^{V N}-T_{4 {\cal E}}^{V N}   \right)
$ \\ \hline
 $N_\alpha^{(6)}$  &  $
-\left(V^p-A^p\right)
   \left(V_{2 n}^{V N}+A_{2 n}^{V N} \right)
   +
   2 T^{p}  \left( T_{2 n}^{V N}+T_{3 n}^{V N}   \right)
$ \\
\end{tabular} \\
\hline
\hline
2 & \raisebox{-0.0cm}
{\includegraphics[height=1.5cm,clip=true]{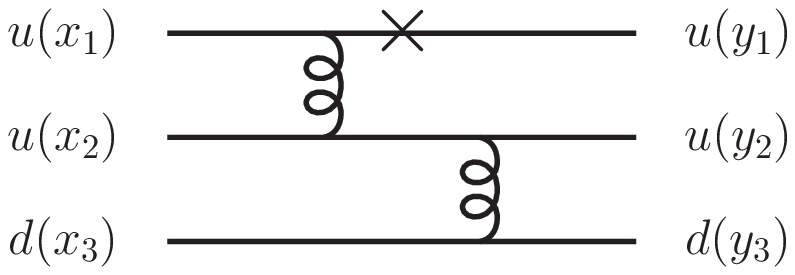}}
 &
\begin{tabular}{p{0.85cm}|p{9.2cm}}
 $N_\alpha^{(1)}$ &  \hspace{4.0cm}    $0$ \\ \hline
 $N_\alpha^{(2)}$ &  \hspace{4.0cm}    $0$ \\ \hline
 $N_\alpha^{(3)}$ &  \hspace{4.0cm}    $0$ \\ \hline
 $N_\alpha^{(4)}$ &  \hspace{4.0cm}    $0$ \\ \hline
 $N_\alpha^{(5)}$ &  \hspace{4.0cm}    $0$ \\ \hline
 $N_\alpha^{(6)}$ &  \hspace{4.0cm}    $0$ \\ 
\end{tabular} \\
\hline
\hline
3 & \raisebox{-0.0cm}
{\includegraphics[height=1.5cm,clip=true]{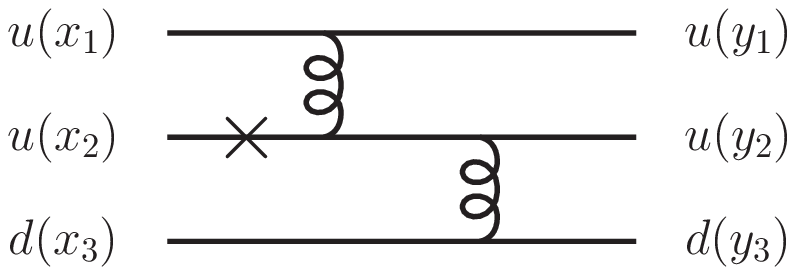}}
$\frac{Q_u (2 \xi)^2)}{(x_{1}-i\epsilon) (2\xi-x_{2}-i\epsilon)(x_{3}-i\epsilon)y_{1}(1-y_{1})y_{3}}$
 &
\begin{tabular}{p{0.85cm}|p{9.2cm}}
 $N_\alpha^{(1)}$ &  $-2 T^{p} \left(T_{1 {\cal E}}^{V N}+T_{2 {\cal E}}^{V N}\right)$ \\ \hline
 $N_\alpha^{(2)}$ & $-4 T^{p} \left(T_{1 n}^{V N}+\frac{\Delta_T^2}{2M^2}T_{4 n}^{V N}\right)$ \\ \hline
 $N_\alpha^{(3)}$ & $-4 T^{p} \left(T_{1 T}^{V N}+T_{3 {\cal E}}^{V N}+\frac{\Delta_T^2}{2M^2}T_{4 T}^{V N}\right)$ \\ \hline
 $N_\alpha^{(4)}$ & $-2 T^{p} \left(T_{2 T}^{V N}+T_{3 T}^{\omega N}\right)$ \\ \hline
 $N_\alpha^{(5)}$ & $2 T^{p} \left(T_{3 {\cal E}}^{V N}-T_{4 {\cal E}}^{V N}\right)$ \\ \hline
 $N_\alpha^{(6)}$ & $2 T^{p} \left(T_{2 n}^{V N}+T_{3 n}^{V N}\right)$ \\ 
\end{tabular} \\
\hline
\hline
4 &  \raisebox{-0.0cm}
{\includegraphics[height=1.5cm,clip=true]{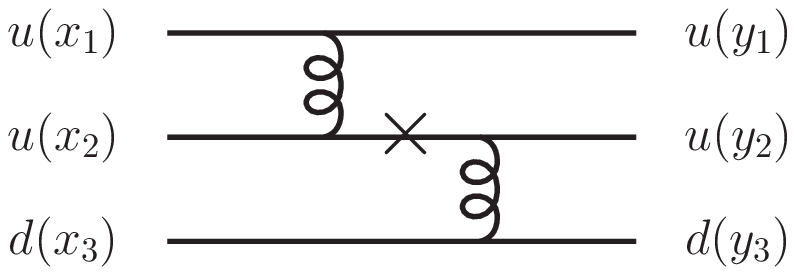}}
$\frac{Q_u (2\xi)^2}{(x_{1}-i\epsilon)
(2\xi-x_{3}-i\epsilon)(x_{3}-i\epsilon)y_{1}(1-y_{1})y_{3}}$
 &
\begin{tabular}{p{0.85cm}|p{9.2cm}}
 $N_\alpha^{(1)}$ &  $-\left(V^{p}-A^{p}\right) \left(V_{1 {\cal E} }^{V N}+A_{1 {\cal E}}^{V N}\right)$ \\ \hline
 $N_\alpha^{(2)}$ & $-\left(V^{p}-A^{p}\right) \left(V_{1 n }^{V N}+A_{1 n}^{V N}\right)$ \\ \hline
 $N_\alpha^{(3)}$ & $-\left(V^{p}-A^{p}\right) \left(V_{1 T }^{V N}+A_{1 T}^{V N}+ V_{2 {\cal E} }^{V N}+A_{2 {\cal E}}^{V N}\right)$ \\ \hline
 $N_\alpha^{(4)}$ & $-\left(V^{p}-A^{p}\right) \left(V_{2 T }^{V N}+A_{2 T}^{V N}\right)$ \\ \hline
 $N_\alpha^{(5)}$ & $\left(V^{p}-A^{p}\right) \left(V_{2 {\cal E} }^{V N}+A_{2 {\cal E}}^{V N}\right)$ \\ \hline
 $N_\alpha^{(6)}$ & $\left(V^{p}-A^{p}\right) \left(V_{2 n }^{V N}+A_{2 n}^{V N}\right)$  \\ 
\end{tabular} \\
\hline
\hline
5 & \raisebox{-0.0cm}
{\includegraphics[height=1.5cm,clip=true]{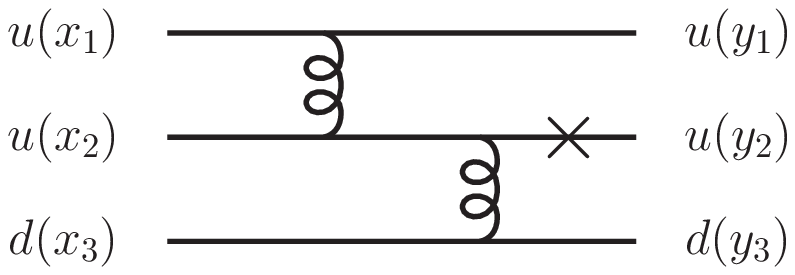}}
$\frac{Q_u (2\xi)^2}{(x_{1}-i\epsilon)(2\xi-x_{3}-i\epsilon)(x_{3}-i\epsilon)y_{1}(1-y_{2})y_{3}}$
 &
\begin{tabular}{p{0.85cm}|p{9.2cm}}
 $N_\alpha^{(1)}$ &  $\left(V^{p}+A^{p}\right) \left(V_{1 {\cal E} }^{V N}-A_{1 {\cal E}}^{V N}\right)$ \\ \hline
 $N_\alpha^{(2)}$ & $\left(V^{p}+A^{p}\right) \left(V_{1 n }^{V N}-A_{1 n}^{V N}\right)$ \\ \hline
 $N_\alpha^{(3)}$ & $\left(V^{p}+A^{p}\right) \left(V_{1T }^{V N}-A_{1 T}+V_{2 {\cal E} }^{V N}-A_{2 {\cal E}}^{V N}\right)$  \\ \hline
 $N_\alpha^{(4)}$ & $\left(V^{p}+A^{p}\right) \left(V_{2T }^{V N}-A_{2 T} \right)$ \\ \hline
 $N_\alpha^{(5)}$ & $-\left(V^{p}+A^{p}\right) \left(V_{2 {\cal E} }^{V N}-A_{2 {\cal E}}^{V N}\right)$ \\ \hline
 $N_\alpha^{(6)}$ & $-\left(V^{p}+A^{p}\right) \left(V_{2 n }^{V N}-A_{2 n}^{V N}\right)$ \\ 
\end{tabular} \\
\hline
\hline
6 & \raisebox{-0.0cm}
{\includegraphics[height=1.5cm,clip=true]{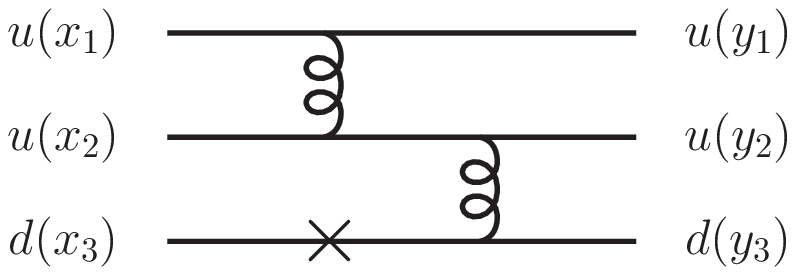}}
 &
\begin{tabular}{p{0.85cm}|p{9.2cm}}
 $N_\alpha^{(1)}$ &  \hspace{4.0cm} $0$ \\ \hline
 $N_\alpha^{(2)}$ & \hspace{4.0cm} $0$ \\ \hline
 $N_\alpha^{(3)}$ & \hspace{4.0cm} $0$ \\ \hline
 $N_\alpha^{(4)}$ & \hspace{4.0cm} $0$ \\ \hline
 $N_\alpha^{(5)}$ & \hspace{4.0cm} $0$ \\ \hline
 $N_\alpha^{(6)}$ & \hspace{4.0cm} $0$ \\ 
\end{tabular} \\
\hline
\hline
7 & \raisebox{-0.0cm}
{\includegraphics[height=1.5cm,clip=true]{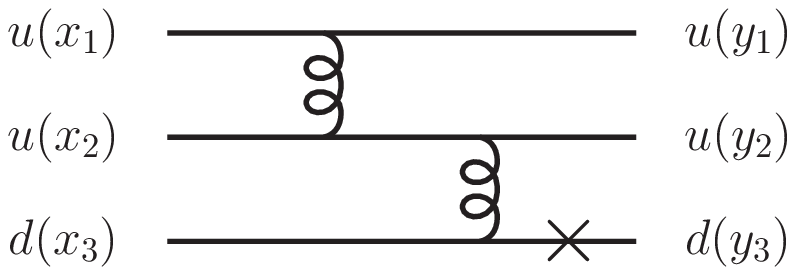}}
$\frac{Q_d(2\xi)^2}{(x_{1}-i\epsilon)(2\xi-x_{3}-i\epsilon)^2y_{1}(1-y_3)^2}$
 &
\begin{tabular}{p{0.85cm}|p{9.2cm}}
 $N_\alpha^{(1)}$ &    $-2 \left(V^{p} V_{1 {\cal E}}^{V N}-A^{p}  A_{1 {\cal E}}^{V N} \right)$ \\ \hline
 $N_\alpha^{(2)}$ & $-2 \left(V^{p} V_{1 n}^{V N}-A^{p}  A_{1 n}^{V N} \right)$ \\ \hline
 $N_\alpha^{(3)}$ &  $-2 \left(V^{p} (V_{1T}^{VN}+V_{2 {\cal E}}^{V N})-A^{p}  (A_{1T}^{VN}+A_{2 {\cal E}}^{V N}) \right)$ \\ \hline
 $N_\alpha^{(4)}$ & $-2 \left(V^{p} V_{2 T}^{V N}-A^{p}  A_{2T}^{V N} \right)$ \\ \hline
 $N_\alpha^{(5)}$ & $2 \left(V^{p} V_{2 {\cal E}}^{V N}-A^{p}  A_{2 {\cal E}}^{V N} \right)$  \\ \hline
 $N_\alpha^{(6)}$ &  $2 \left(V^{p} V_{2 n}^{V N}-A^{p}  A_{2 n}^{V N} \right)$\\ 
\end{tabular} \\
\hline
\hline
8 & \raisebox{-0.0cm}
{\includegraphics[height=1.5cm,clip=true]{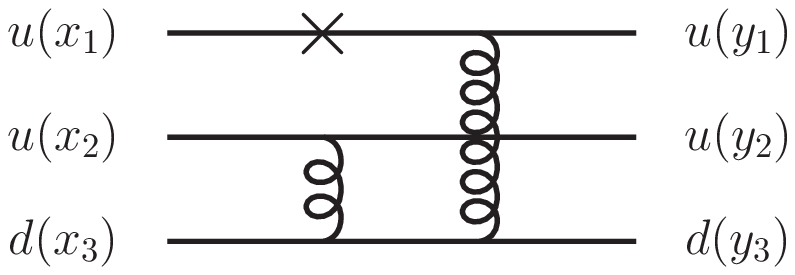}}
 &
\begin{tabular}{p{0.85cm}|p{9.2cm}}
 $N_\alpha^{(1)}$ &  \hspace{4.0cm} $0$ \\ \hline
 $N_\alpha^{(2)}$ & \hspace{4.0cm} $0$ \\ \hline
 $N_\alpha^{(3)}$ & \hspace{4.0cm} $0$ \\ \hline
 $N_\alpha^{(4)}$ & \hspace{4.0cm} $0$ \\ \hline
 $N_\alpha^{(5)}$ & \hspace{4.0cm} $0$ \\ \hline
 $N_\alpha^{(6)}$ & \hspace{4.0cm} $0$ \\ 
\end{tabular} \\
\hline
\hline
9 & \raisebox{-0.0cm}
{\includegraphics[height=1.5cm,clip=true]{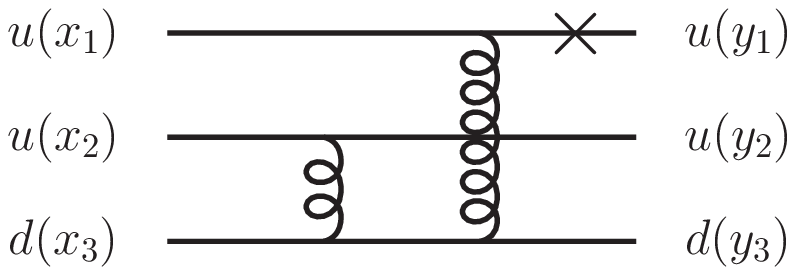}}
 \ \ \ \ \ \ \ \  \ \ \ \ \
$\frac{Q_u (2\xi)^2}{(2\xi-x_{1}-i\epsilon)^2(x_{2}-i\epsilon)(1-y_{1})^2y_{2}}$ &
\begin{tabular}{p{0.85cm}|p{9.2cm}}
 $N_\alpha^{(1)}$ & $
-\left(V^p-A^p\right)
   \left(V_{1 {\cal E}}^{V N}+A_{1 {\cal E}}^{V N}\right)
   +
   2 T^{p} \left(T_{1 {\cal E}}^{V N}+T_{2 {\cal E}}^{V N}\right)
$ \\ \hline
 $N_\alpha^{(2)}$
 & $ -\left(V^p-A^p\right)
   \left(V_{1 n}^{V N}+A_{1 n}^{V N}\right)
   +
   4 T^{p}  \left( T_{1 n}^{V N} + \frac{\Delta_T^2 }{2 M^2} T_{4 n}^{V N} \right)
$
\\ \hline
\raisebox{-0.2cm}{  $N_\alpha^{(3)}$ } &
$
-\left(V^p-A^p\right)
   \left(V_{1 T}^{V N}+A_{1 T}^{V N}+V_{2 {\cal E}}^{V N}+A_{2 {\cal E}}^{V N}\right)
  $
\\
    &  $ +
   4 T^{p}  \left( T_{1 T}^{V N}+T_{3 {\cal E}}^{V N} + \frac{\Delta_T^2  }{2 M^2} T_{4T}^{V N} \right)$  \\ \hline
  $N_\alpha^{(4)}$ &  $
-\left(V^p-A^p\right)
   \left(V_{2 T}^{V N}+A_{2 T}^{V N} \right)
   +
   2 T^{p}  \left( T_{2 T}^{V N}+T_{3 T}^{V N}   \right)
$ \\ \hline
 $N_\alpha^{(5)}$  &  $
\left(V^p-A^p\right)
   \left(V_{2 {\cal E}}^{V N}+A_{2 {\cal E}}^{V N} \right)
   -
   2 T^{p}  \left( T_{3 {\cal E}}^{V N}-T_{4 {\cal E}}^{V N}   \right)
$ \\ \hline
 $N_\alpha^{(6)}$  &  $
-\left(V^p-A^p\right)
   \left(V_{2 n}^{V N}+A_{2 n}^{V N} \right)
   +
   2 T^{p}  \left( T_{2 n}^{V N}+T_{3 n}^{V N}   \right)
$ \\
\end{tabular} \\
\hline
\hline
10 & \raisebox{-0.0cm}
{\includegraphics[height=1.5cm,clip=true]{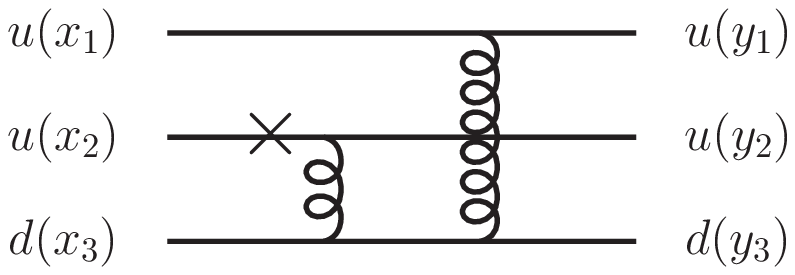}}
 \ \ \ \ \ \ \ \  \ \ \ \ \
$\frac{Q_u (2\xi)^2}{(x_{1}-i\epsilon)(2\xi-x_{2}-i\epsilon)^2y_{1}(1-y_{2})^2}$ &
\begin{tabular}{p{0.85cm}|p{9.2cm}}
 $N_\alpha^{(1)}$ & $
-\left(V^p-A^p\right)
   \left(V_{1 {\cal E}}^{V N}+A_{1 {\cal E}}^{V N}\right)
   +
   2 T^{p} \left(T_{1 {\cal E}}^{V N}+T_{2 {\cal E}}^{V N}\right)
$ \\ \hline
 $N_\alpha^{(2)}$
 & $ -\left(V^p-A^p\right)
   \left(V_{1 n}^{V N}+A_{1 n}^{V N}\right)
   +
   4 T^{p}  \left( T_{1 n}^{V N} + \frac{\Delta_T^2 }{2 M^2} T_{4 n}^{V N} \right)
$
\\ \hline
\raisebox{-0.2cm}{  $N_\alpha^{(3)}$ } &
$
-\left(V^p-A^p\right)
   \left(V_{1 T}^{V N}+A_{1 T}^{V N}+V_{2 {\cal E}}^{V N}+A_{2 {\cal E}}^{V N}\right)
  $
\\
    &  $ +
   4 T^{p}  \left( T_{1 T}^{V N}+T_{3 {\cal E}}^{V N} + \frac{\Delta_T^2  }{2 M^2} T_{4T}^{V N} \right)$  \\ \hline
  $N_\alpha^{(4)}$ &  $
-\left(V^p-A^p\right)
   \left(V_{2 T}^{V N}+A_{2 T}^{V N} \right)
   +
   2 T^{p}  \left( T_{2 T}^{V N}+T_{3 T}^{V N}   \right)
$ \\ \hline
 $N_\alpha^{(5)}$  &  $
\left(V^p-A^p\right)
   \left(V_{2 {\cal E}}^{V N}+A_{2 {\cal E}}^{V N} \right)
   -
   2 T^{p}  \left( T_{3 {\cal E}}^{V N}-T_{4 {\cal E}}^{V N}   \right)
$ \\ \hline
 $N_\alpha^{(6)}$  &  $
-\left(V^p-A^p\right)
   \left(V_{2 n}^{V N}+A_{2 n}^{V N} \right)
   +
   2 T^{p}  \left( T_{2 n}^{V N}+T_{3 n}^{V N}   \right)
$ \\
\end{tabular} \\
\hline
\hline
11 & \raisebox{-0.0cm}
{\includegraphics[height=1.5cm,clip=true]{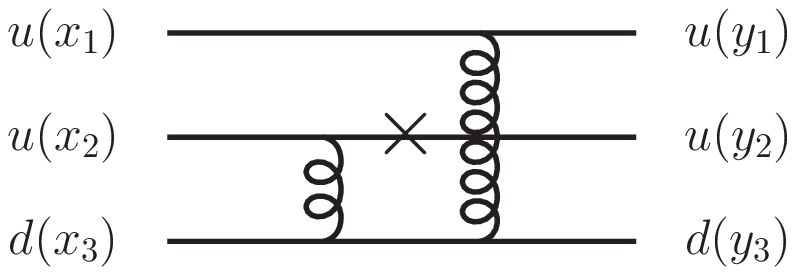}}
 &
\begin{tabular}{p{0.85cm}|p{9.2cm}}
 $N_\alpha^{(1)}$ &  \hspace{4.0cm} $0$ \\ \hline
 $N_\alpha^{(2)}$ & \hspace{4.0cm} $0$ \\ \hline
 $N_\alpha^{(3)}$ & \hspace{4.0cm} $0$ \\ \hline
 $N_\alpha^{(4)}$ & \hspace{4.0cm} $0$ \\ \hline
 $N_\alpha^{(5)}$ & \hspace{4.0cm} $0$ \\ \hline
 $N_\alpha^{(6)}$ & \hspace{4.0cm} $0$ \\ 
\end{tabular} \\
\hline
\hline
12 & \raisebox{-0.0cm}
{\includegraphics[height=1.5cm,clip=true]{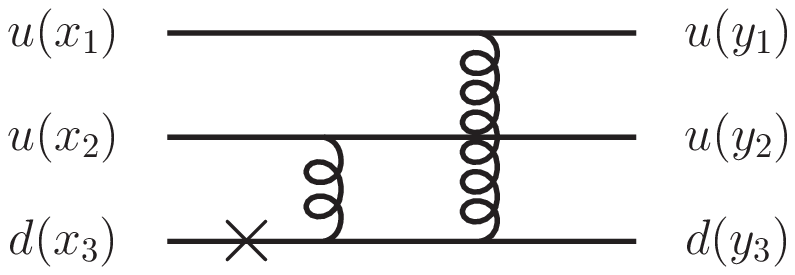}}
$\frac{Q_d (2 \xi)^2}{(x_{1}-i\epsilon)(x_{2}-i\epsilon)(2\xi-x_{3}-i\epsilon)y_{1}(1-y_{2})y_{2}}$
 &
\begin{tabular}{p{0.85cm}|p{9.2cm}}
 $N_\alpha^{(1)}$ &  $\left(V^{p}+A^{p}\right) \left( V_{1 {\cal E}}^{V N}-A_{1 {\cal E}}^{V N}\right)$
 \\ \hline
 $N_\alpha^{(2)}$ &  $\left(V^{p}+A^{p}\right) \left( V_{1 {n}}^{V N}-A_{1 {n}}^{V N}\right)$  \\ \hline
 $N_\alpha^{(3)}$ & $\left(V^{p}+A^{p}\right) \left( V_{1 {T}}^{V N}-A_{1 {T}}^{V N}+V_{2 {\cal E}}^{V N}-A_{2 {\cal E}}^{V N}\right)$  \\ \hline
 $N_\alpha^{(4)}$ & $\left(V^{p}+A^{p}\right) \left( V_{2 {T}}^{V N}-A_{2 {T}}^{V N}\right)$ \\ \hline
 $N_\alpha^{(5)}$ & $-\left(V^{p}+A^{p}\right) \left( V_{2 {\cal E}}^{V N}-A_{2 {\cal E}}^{V N}\right)$  \\ \hline
 $N_\alpha^{(6)}$ & $-\left(V^{p}+A^{p}\right) \left( V_{2 n}^{V N}-A_{2 n}^{V N}\right)$  \\ 
\end{tabular} \\
\hline
\hline
13 & \raisebox{-0.0cm}
{\includegraphics[height=1.5cm,clip=true]{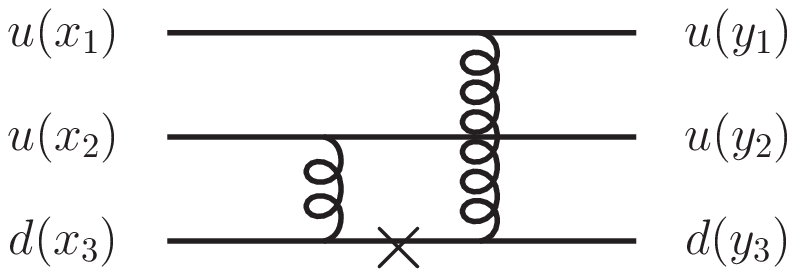}}
$\frac{Q_d (2 \xi)^2)}{(x_{1}-i\epsilon)(2\xi-x_{1}-i\epsilon)(x_{2}-i\epsilon)y_{1}(1-y_{2})y_{2}}$
 &
\begin{tabular}{p{0.85cm}|p{9.2cm}}
 $N_\alpha^{(1)}$ &  $2 T^{p} \left(T_{1 {\cal E}}^{V N}+T_{2 {\cal E}}^{V N}\right)$ \\ \hline
 $N_\alpha^{(2)}$ & $4 T^{p} \left(T_{1 n}^{V N}+\frac{\Delta_T^2}{2 M^2}T_{4 n}^{V N}\right)$ \\ \hline
 $N_\alpha^{(3)}$ & $4 T^{p} \left(T_{1 T}^{V N}+T_{3 {\cal E}}^{V N}+\frac{\Delta_T^2}{2 M^2}T_{4 T}^{V N}\right)$ \\ \hline
 $N_\alpha^{(4)}$ & $2 T^{p} \left(T_{2 T}^{V N}+T_{3 T}^{\omega N}\right)$ \\ \hline
 $N_\alpha^{(5)}$ & $-2 T^{p} \left(T_{3 {\cal E}}^{V N}-T_{4 {\cal E}}^{V N}\right)$ \\ \hline
 $N_\alpha^{(6)}$ & $-2 T^{p} \left(T_{2 n}^{V N}+T_{3 n}^{V N}\right)$ \\ 
\end{tabular} \\
\hline
\hline
14 &  \raisebox{-0.0cm}
{\includegraphics[height=1.5cm,clip=true]{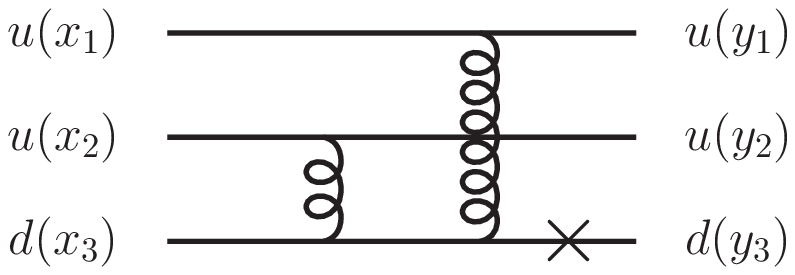}}
$\frac{Q_d (2\xi)^2}{(x_{1}-i\epsilon)(2\xi-x_{1}-i\epsilon)(x_{2}-i\epsilon)y_{1}y_{2}(1-y_{3})}$
 &
\begin{tabular}{p{0.85cm}|p{9.2cm}}
 $N_\alpha^{(1)}$ &  $\left(V^{p}-A^{p}\right) \left(V_{1 {\cal E} }^{V N}+A_{1 {\cal E}}^{V N}\right)$ \\ \hline
 $N_\alpha^{(2)}$ & $\left(V^{p}-A^{p}\right) \left(V_{1 n }^{V N}+A_{1 n}^{V N}\right)$ \\ \hline
 $N_\alpha^{(3)}$ & $\left(V^{p}-A^{p}\right) \left(V_{1 T }^{V N}+A_{1 T}^{V N}+ V_{2 {\cal E} }^{V N}+A_{2 {\cal E}}^{V N}\right)$ \\ \hline
 $N_\alpha^{(4)}$ & $\left(V^{p}-A^{p}\right) \left(V_{2 T }^{V N}+A_{2 T}^{V N}\right)$ \\ \hline
 $N_\alpha^{(5)}$ & $-\left(V^{p}-A^{p}\right) \left(V_{2 {\cal E} }^{V N}+A_{2 {\cal E}}^{V N}\right)$ \\ \hline
 $N_\alpha^{(6)}$ & $-\left(V^{p}-A^{p}\right) \left(V_{2 n }^{V N}+A_{2 n}^{V N}\right)$  \\ 
\end{tabular} \\
\hline
\end{longtable}


\section{Calculation of the unpolarized cross section}
\label{Sec_CS}
First of all, we need to specify our conventions for the backward
vector meson electroproduction cross section. Within the suggested
factorization mechanism only the transverse virtual photoproduction cross section
$\sigma_T$
receives
contribution at the leading twist.
Analogously to how this was done in Ref.~\cite{Lansberg:2011aa}, using the explicit expression
(eq. (2.12) of \cite{Kroll:1995pv})
relating scattering amplitudes of vector meson electroproduction
within one photon approximation and the amplitudes for the virtual photoproduction,
we express the unpolarized cross section of hard electroproduction of
a $V$ meson off a nucleon
through the helicity amplitudes of
$\gamma^* N \to N V$
defined in
(\ref{Hel_ampl_def}) within the collinear factorized description framework:
\be
&&
\frac{d^5 \sigma}{dE' d \Omega_{e'} d \Omega_V}
\nonumber \\ &&
= \Gamma \times \frac{\Lambda(s,m_V^2,M^2)}{128 \pi^2 s (s-M^2)}
\left\{  \frac{1}{2} \sum_{{\lambda_\gamma}_T, \, \lambda_V, \, s_1, \, s_2 } |{\cal M}_{s_1 s_2}^{\lambda_\gamma \lambda_V}|^2 + \ldots\right\}=
\Gamma \times \left\{  \frac{d^2 \sigma_T}{d \Omega_V} + \ldots \right\}.
\label{CS_def_formula}
\ee
Here
$\Omega_{e'}$
is the differential solid angle for the scattered electron in the LAB frame;
$\Omega_V$
is  the differential solid angle of the produced vector meson in
$N' V$
CMS frame;
and $\Lambda$ is the Mandelstam function
(\ref{Def_lambda}).

By dots in the r.h.s. of eq.~(\ref{CS_def_formula}) we denote the subleading twist terms supressed by powers of $1/Q$.
$\Gamma$ stands for the virtual photon flux factor in Hand's convention
\be
\Gamma = \frac{\alpha_{\rm em}}{2 \pi^2} \frac{{k'}_{0 }^L}{k_{0 }^L} \frac{s-M^2}{2 M Q^2} \frac{1}{1-\epsilon}.
\ee
Here
$k_0^L$
and
$k_0^{'L}$ denote the initial state and final state electron energies in the LAB frame and
\be
\epsilon=\Big[1+2\frac{\big({k}_0^L-{k'}_0^L \big)^2+Q^2}{Q^2} \tan^2 \frac{\theta^L_e }{2}\Big]^{-1},
\ee
is the polarization parameter of the virtual photon,
where
$\theta^L_e$
denotes the electron scattering angle in the LAB frame.

We employ the following relation for the sum over photon's transverse polarizations
\be
\sum_{{\lambda_\gamma}_T} {\varepsilon}^\nu (q,\lambda_\gamma) {\varepsilon}^{\mu *} (q,\lambda_\gamma)=- g^{\mu \nu}+ \frac{1}{(p \cdot n)} (p^\mu n^\nu +p^\nu n^\mu),
\ee
and the $V$-meson polarization sum reads
\be
\sum_{\lambda_V }{\cal E}^\rho (p_V,\lambda_V) {\cal E}^\sigma (p_V,\lambda_V)=- g^{\rho \sigma}+ \frac{p_V^\rho p_V^\sigma}{m_V^2}.
\ee

Let us introduce the notation
\be
{\cal D}=-i \left( \frac{(4 \pi \alpha_s)^2 \sqrt{4 \pi \alpha_{em}} f_N M}{54} \right),
\ee
and define
\be
&&
\Gamma_H  =
\hat{\varepsilon}(q, \lambda_\gamma) \hat{\cal E}(p_V, \lambda_V)
{\cal I}^{(1)}(\xi, \, \Delta^2)+
M ( {\cal E} (p_V, \lambda_V) \cdot n)\hat{\varepsilon}(q, \lambda_\gamma)
\,
{\cal I}^{(2)}(\xi, \, \Delta^2)
\nonumber \\ &&
+\frac{({\cal E}(p_V, \lambda_V) \cdot \Delta_T)}{M}\hat{\varepsilon}(q, \lambda_\gamma)
\,
{\cal I}^{(3)}(\xi, \, \Delta^2)
+\frac{({\cal E}(p_V, \lambda_V) \cdot \Delta_T)}{M^2}\hat{\varepsilon}(q, \lambda_\gamma) \hat{\Delta}_T
\,
{\cal I}^{(4)}(\xi, \, \Delta^2)
\nonumber \\ &&
+\frac{1}{M} \hat{\varepsilon}(q, \lambda_\gamma)  \hat{\cal E}(p_V, \lambda_V) \hat{\Delta}_T
\,
{\cal I}^{(5)}(\xi, \, \Delta^2)+
( {\cal E} (p_V, \lambda_V) \cdot n) \hat{\cal E}(p_V, \lambda_V) \hat{\Delta}_T {\cal I}^{(6)}(\xi, \, \Delta^2)\,.
\ee
We now square the amplitude and sum over the transverse polarization of the virtual photon,
over the spin of outgoing nucleon and over the polarizations of the $V$-meson and sum over
the spin of the initial nucleon.
We make use of kinematic relations summarized in
(\ref{Sudakov_dec_omega})
and keep only the leading twist terms.
The resulting expression then reads:
\be
&&
|{\cal M}_T|^2=   \sum_{{\lambda_\gamma}_T, \, \lambda_V, \, s_1, \, s_2 } {\cal M}_{s_1 s_2}^{\lambda_\gamma \lambda_V} {{\cal M}_{s_1 s_2}^{\lambda_\gamma \lambda_V}}^*
=|{\cal D}|^2 \frac{1}{Q^8}
\sum_{{\lambda_\gamma}_T, \, \lambda_\omega}   {\rm Tr}
\left\{
(\hat{p}_2+M) \Gamma_H (\hat{p}_1+M) \gamma_0 \Gamma_H^\dag \gamma_0
\right\} \nonumber \\
&&
= |{\cal D}|^2 \frac{1}{Q^6} \frac{2(1+\xi)}{\xi}  \left[  |{\cal I}^{(1)} |^2
\left( 1+
\frac{ M^2(1-\xi)^2+(1+\xi)^2(m_V^2-\Delta_T^2)}{ m_V^2(1+\xi)^2}
\right)
   \right.
\nonumber \\
&&
\left.
+ |{\cal I}^{(2)}  |^2 \frac{M^2 (1-\xi)^2   }{4 m_V^2  } +
\left(
{\cal I}^{(1)}   {{\cal I}^{(2)*}}   +
{\cal I}^{(1)*}   {{\cal I}^{(2)}}
\right) \frac{M^2 (1-\xi)^2}{2 m_V^2 (1+\xi)}\right.
\nonumber \\ && \left. +\frac{\Delta_T^2}{M^2} \left\{
-|{\cal I}^{(3)}  |^2\frac{ \left(m_V^2-\Delta_T^2\right)}{
   m_V^2  } +
\left(
{\cal I}^{(1)}   {{\cal I}^{(3)*}}   +
{\cal I}^{(1)*}   {{\cal I}^{(3)}}
\right)
\frac{M^2 (1-\xi)}{ m_V^2 (1+ \xi) }
\right.
\right.
\nonumber  \\ &&
\left.
\left.
 +
\left(
{\cal I}^{(2)}   {{\cal I}^{(3)*}}   +
{\cal I}^{(2)*}   {{\cal I}^{(3)}}
\right)
\frac{M^2(1-\xi)}{2 m_V^2 }+
|{\cal I}^{(4)}|^2
\frac{\Delta_T^2(m_V^2-\Delta_T^2)}{M^2 m_V^2}
\right.
\right.
\nonumber  \\ &&
\left.
\left.
+\left(
{\cal I}^{(1)}   {{\cal I}^{(4)*}}   +
{\cal I}^{(1)*}   {{\cal I}^{(4)}}
\right)
\frac{ (m_V^2-\Delta_T^2)}{ m_V^2}-
|{\cal I}^{(5)}  |^2
\left( 1+
\frac{ M^2(1-\xi)^2+(1+\xi)^2(m_V^2-\Delta_T^2)}{ m_V^2(1+\xi)^2}
\right)
\right.
\right.
\nonumber  \\ &&
\left.
\left.
+\left(
{\cal I}^{(1)}   {{\cal I}^{(5)*}}   +
{\cal I}^{(1)*}   {{\cal I}^{(5)}}
\right) \frac{2M^2(1-\xi)}{m_V^2(1+\xi)}
+\left(
{\cal I}^{(2)}   {{\cal I}^{(5)*}}   +
{\cal I}^{(2)*}   {{\cal I}^{(5)}}
\right)
\frac{M^2(1-\xi)}{2 m_V^2}
\right.
\right.
\nonumber  \\ &&
\left.
\left.
-\left(
{\cal I}^{(3)}   {{\cal I}^{(5)*}}   +
{\cal I}^{(3)*}   {{\cal I}^{(5)}}
\right)
\frac{m_V^2-\Delta_T^2}{m_V^2}+
\left(
{\cal I}^{(4)}   {{\cal I}^{(5)*}}   +
{\cal I}^{(4)*}   {{\cal I}^{(5)}}
\right)
\frac{\Delta_T^2(1-\xi)}{m_V^2(1+\xi)}
\right.
\right.
\nonumber  \\ &&
\left.
\left.
-|{\cal I}^{(6)}  |^2 \frac{M^2(1-\xi)^2}{4m_V^2}-
\left(
{\cal I}^{(1)}   {{\cal I}^{(6)*}}   +
{\cal I}^{(1)*}   {{\cal I}^{(6)}}
\right)
\frac{M^2(1-\xi)}{2 m_V^2}
\right.
\right.
\nonumber  \\ &&
\left.
\left.
-\left(
{\cal I}^{(4)}   {{\cal I}^{(6)*}}   +
{\cal I}^{(4)*}   {{\cal I}^{(6)}}
\right)
\frac{\Delta_T^2(1-\xi)}{2 m_V^2}+
\left(
{\cal I}^{(5)}   {{\cal I}^{(6)*}}   +
{\cal I}^{(5)*}   {{\cal I}^{(6)}}
\right)
\frac{M^2(1-\xi)^2}{2 m_V^2(1+\xi)}
\right\}
\right].
\ee

We end up with the following expression for the LO unpolarized cross section
of hard photoproduction of backward vector mesons off nucleons:
\be
&&
\frac{d^2 \sigma_T}{d \Omega_V}=  \frac{\Lambda(s,m_V^2,M^2)}{128 \pi^2 s (s-M^2)}  \frac{1}{2} |{\cal M}_T|^2,
\label{CS_formula_def}
\ee
where
$\sigma_T$
refers to the transverse polarization of the virtual photon and
$\frac{1}{2}$ stands for averaging over the initial nucleon spin.

Thus we conclude that there are two essential marking signs of the onset of the suggested
factorization regime for hard vector meson production in the near-backward kinematics,
which can be tested experimentally.
\bi
\item The dominance of the transverse polarization of the virtual photon resulting in the
suppression of the
$\sigma_L$
cross section by at least
$1/Q^2$.
In fact, the preliminary analysis
\cite{HuberPrivate}
of backward $\omega$-meson production JLab Hall C $6$ GeV data hints at
$\sigma_T > \sigma_L$
already for
$Q^2\simeq2.4$ GeV$^2$
and
$W\simeq2.2$ GeV.
\item The characteristic $1/Q^8$-scaling behavior of the transverse cross section (\ref{CS_formula_def}) for fixed
$x_{B}$.
\ei

In what follows we present our estimates for the LO unpolarized cross section
of hard photoproduction of backward
$\omega(782)$, $\phi(1020)$ and $\rho^0(770)$
mesons off protons within the $u$-channel nucleon
exchange model for $VN$ TDAs presented in the Appendix~\ref{App_NPole}.
This is a simple TDA model which populates $VN$ TDAs only within the ERBL-like support region.
However, it can be seen as a reliable estimate of the $VN$ TDAs magnitude for the intermediate values
of skewness parameter
$\xi=0.1 \div 0.4$.
As inputs this model requires the values of the vector and tensor 
$G_{VNN}^{V, \, T}$ 
couplings 
(\ref{GTGV_couplings})
and the phenomenological solutions for the leading twist nucleon DAs.

\subsection*{On the choice of phenomenological parametrization for the nucleon DA}

The choice of the phenomenological solution for the leading twist nucleon DA and the corresponding
value of the strong coupling represents a complicated problem
(see {\it e.g.} discussion in Ref.~\cite{Stef}). Roughly speaking, there exist two distinct classes 
of the leading twist nucleon DA parametrizations.
\bi
\item The models with the shape of nucleon DA close to the asymptotic form \cite{Lepage:1980fj}  
\be
V^p(y_1,y_2,y_3)=T^p(y_1,y_2,y_3)=120 y_1 y_2 y_3; \ \ \
A^p(y_1,y_2,y_3)=0
\label{As_f_DA}
\ee
already at a low normalization scale.
Prominent examples are the Bolz-Kroll (BK) \cite{Bolz:1996sw} and Braun-Lenz-Wittmann (BLW) LO and NLO models
\cite{Braun:2006hz,Lenz:2009ar}. Also, the advanced lattice calculations of the nucleon DA
\cite{Braun:2014wpa}
favor such nucleon DAs. 

\item The Chernyak-Zhitnitsky (CZ)-type models with a shape of nucleon DA
considerably different from the  the asymptotic limit at a low normalization scale. 
The examples of this type of nucleon DA models are the  
CZ \cite{Chernyak:1984bm}, King-Sachrajda (KS) \cite{King:1986wi}, 
Chernyak-Ogloblin-Zhitnitsky (COZ) 
\cite{Chernyak:1987nv}
and Gari-Stefanis (GS) 
\cite{Gari:1986ue} solutions.
\ei

Both types of nucleon DA models were employed to provide a description of the nucleon electromagnetic form factors. 
As it is well known, the asymptotic form of the nucleon DA 
(\ref{As_f_DA})
results in a vanishing pQCD contribution for the proton form factor. Therefore, using the nucleon DA with a shape close to the asymptotic form
implies that the standard pQCD contribution must be 
be complemented by the so-called soft or end-point corrections (see {\it e.g.} discussion in Refs. 
\cite{Radyushkin:1990te,Bolz:1996sw,Braun:2006hz,Anikin:2013aka}).

However,  the nucleon 
electromagnetic form factor appears as a building block of the backward amplitude within our $u$-channel nucleon exchange model for $VN$ TDAs.
Therefore, within our simple model for $VN$ TDAs, we need to assure that the pQCD contribution into
the nucleon electromagnetic form factor is close to the experimental value.
This implies that we are forced to use the CZ-type solutions for nucleon DA.

In the following phenomenological estimates we have chosen to employ the COZ 
\cite{Chernyak:1987nv} 
and 
KS 
\cite{King:1986wi} solutions for the leading twist nucleon DAs and set
a compromise value of the strong coupling 
$\alpha_s = 0.3$.

\subsection*{Backward $\omega$ -meson hard photoproduction}

As the first example we consider
backward $\omega(782)$ meson hard photoproduction off proton
\be
\gamma^*(q,\lambda_\gamma)+ p(p_1,s_1) \to \omega(p_\omega, \lambda_\omega)+p(p_2,s_2).
\ee
We take the{ \verb"Grein'80"} estimates for the $\omega$-meson couplings to nucleons
\cite{Grein:1979nw}
(see also Table 9.2 of Ref.~\cite{Dumbrajs:1983jd})
\be
G^V_{\omega NN}=10.1 ; \ \ \ G^T_{\omega NN}=1.42.
\ee

\begin{figure}[H]
 \begin{center}
 \epsfig{figure= 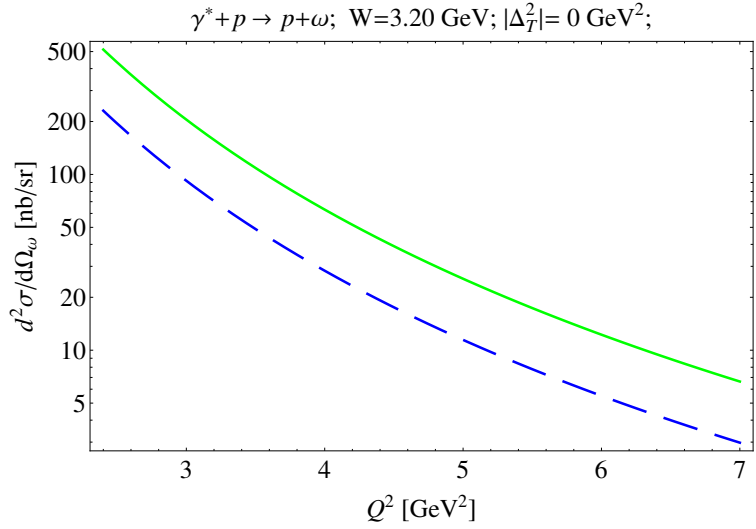 , height=7cm}
  \end{center}
  \caption{
Unpolarized cross section
$\frac{d^2 \sigma_T}{d \Omega_V}$
(in nb/sr) for backward
$\gamma^*  + p  \to p +\omega $
for fixed $W=3.20$ GeV as
a function of $Q^2$ in the $u$-channel nucleon exchange model for $\omega N$ TDAs.
COZ
\cite{Chernyak:1987nv}
(long-dashed lines) and
KS
\cite{King:1986wi}
(solid)
solutions for the leading twist nucleon DA
are used as the phenomenological input.
  }
\label{FigCSomegaQ2}
\end{figure}

We take the kinematical point from the foreseen Jlab@$12$ GeV  setup for Hall C
({\verb"Fpi-3 (E06-12-101)"} $u$-channel kinematics)
\cite{EMPworkshop}
$W=3.20$~GeV.
On Fig.~\ref{FigCSomegaQ2}
we plot the resulting unpolarized cross section
$\frac{d^2 \sigma_T}{d \Omega_\omega}$
(\ref{CS_formula_def})
within the $u$-channel nucleon exchange model for $\omega N$ TDAs
as a function of $Q^2$ for $\Delta_T^2=0$
({\it i.e.} for the $\omega$-meson produced exactly in the backward direction in the
$\gamma^* N$ CMS: $\theta_\omega^*=\pi$).

\begin{figure}[H]
 \begin{center}
  \epsfig{figure= 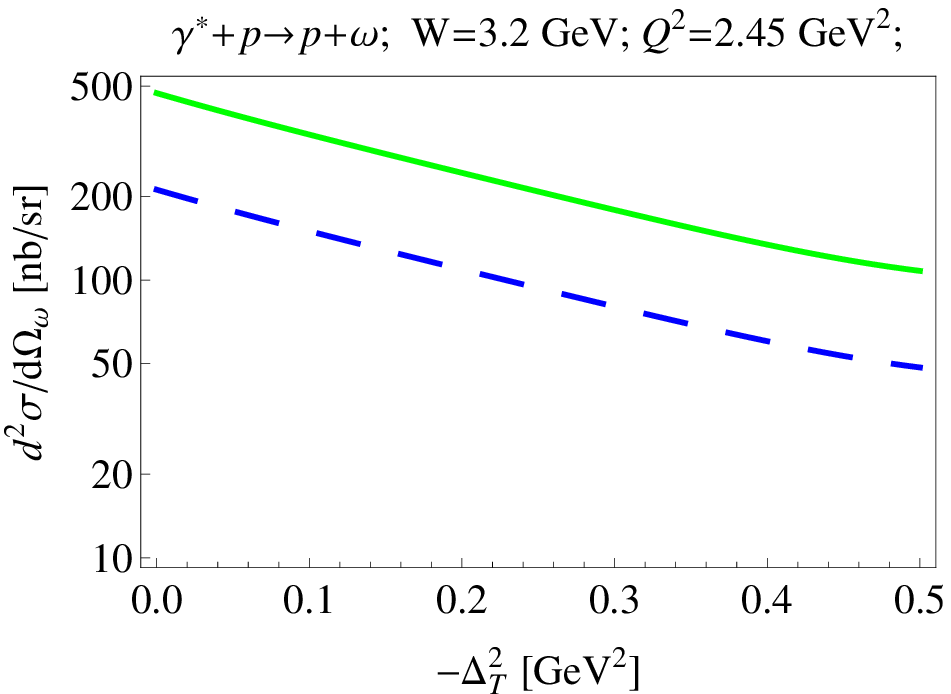 , height=5.5cm}
  \epsfig{figure= 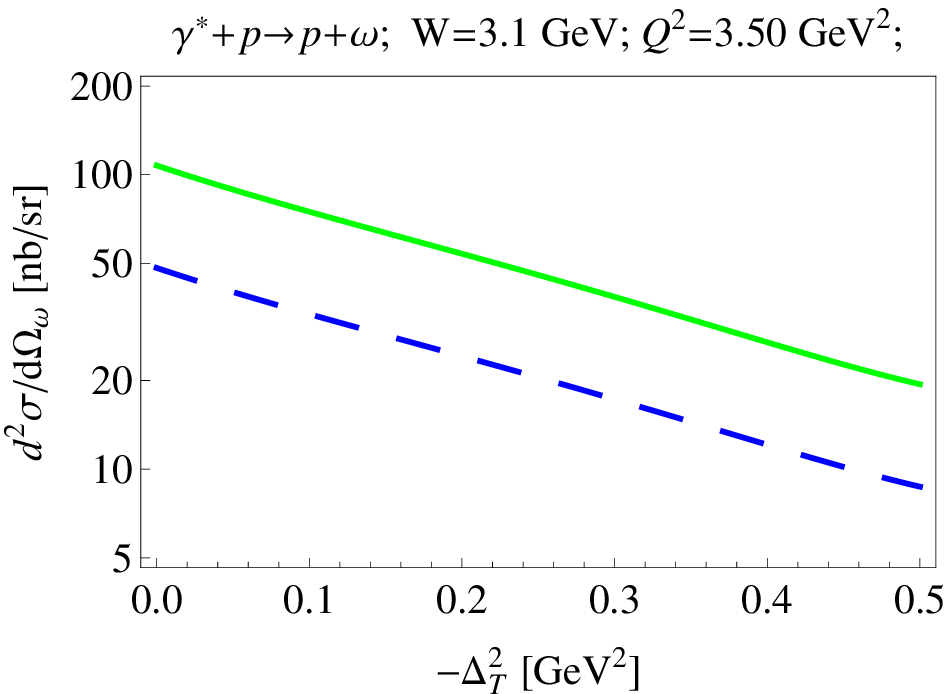 , height=5.5cm}
  \epsfig{figure= 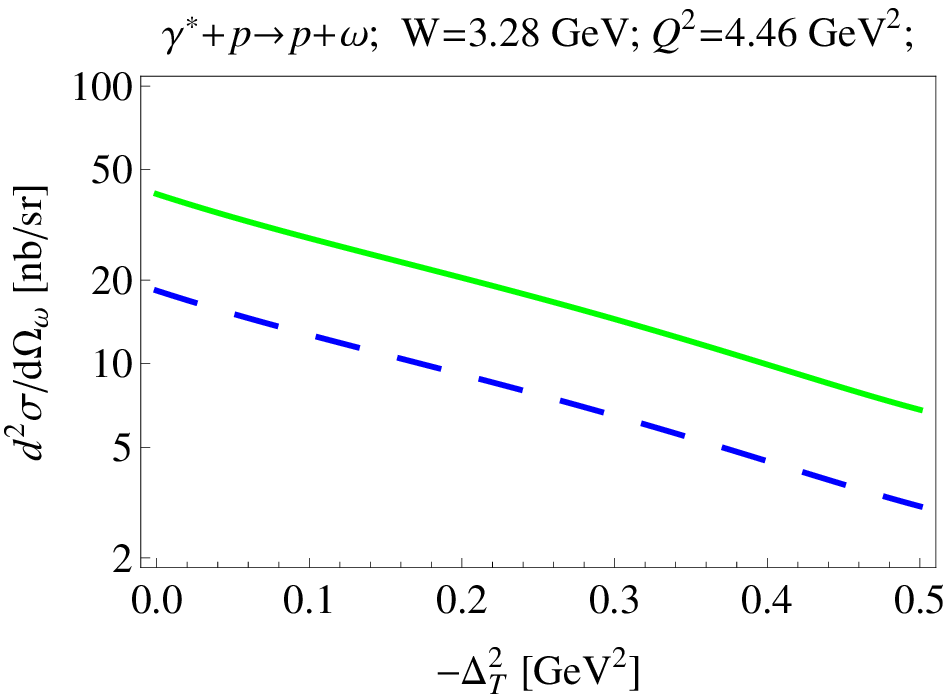 , height=5.5cm}
  \epsfig{figure= 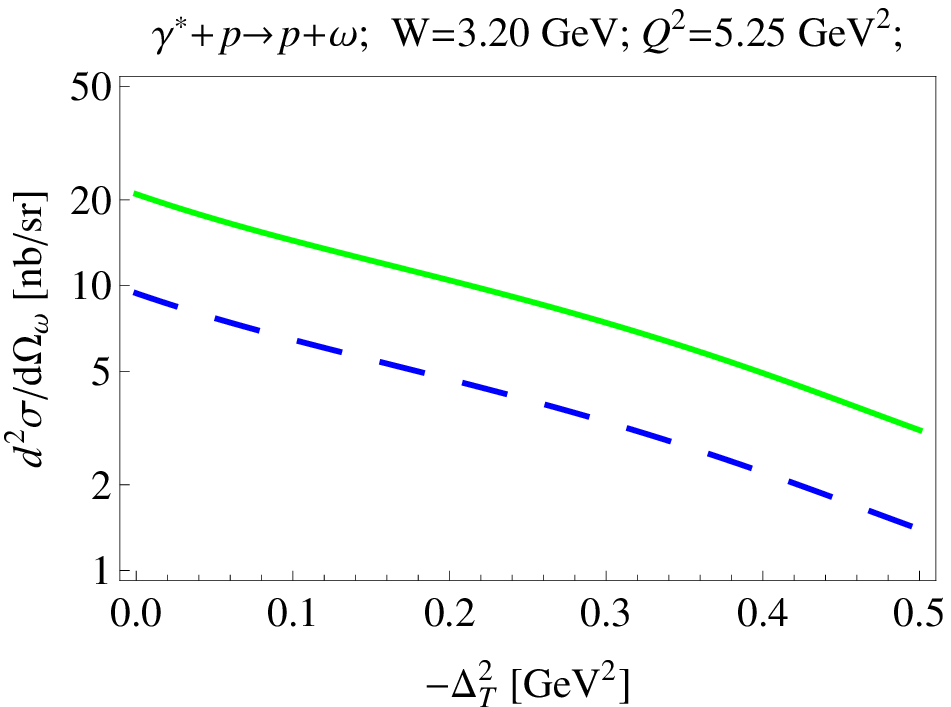 , height=5.5cm}
  \epsfig{figure= 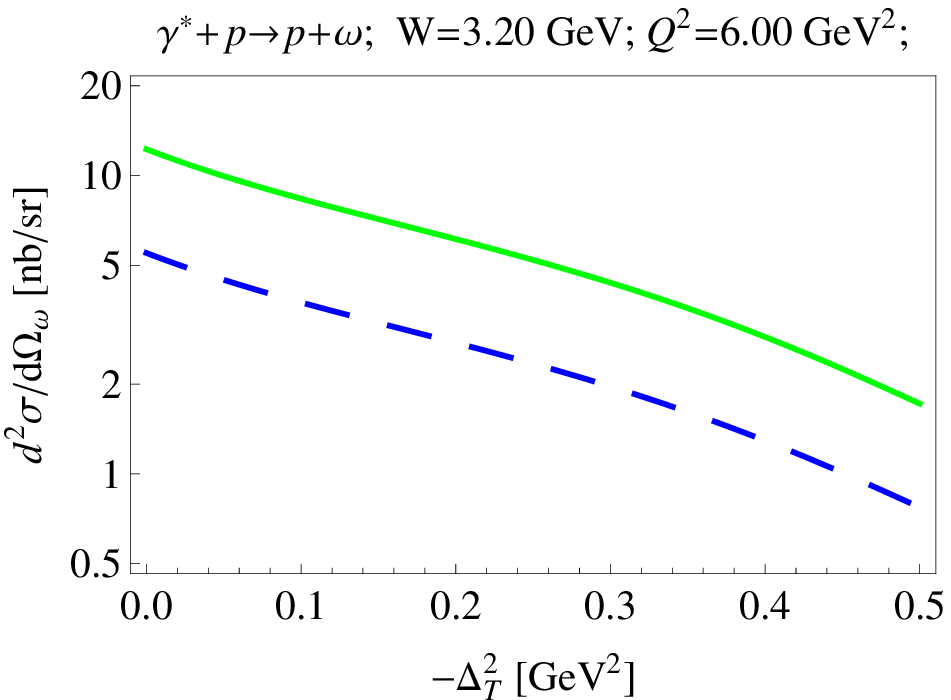 , height=5.5cm}
  \ \ \ \ \ \ \ \ \ \ \ \ \ \ \ \ \ \ \ \ \ \ \ \ \ \ \  \ \ \ \ \ \ \ \ \ \ \ \ \ \ \ \ \ \ \ \ \ \ \ \ \ \ \ \ \ \ \ \ \ \
 \end{center}

 \caption{Unpolarized cross section
$\frac{d^2 \sigma_T}{d \Omega_\omega}$
(in nb/sr) for backward
$\gamma^*  + p  \to p +\omega $
for several values of $W$ and $Q^2$
as a function of $\Delta_T^2$ in the $u$-channel nucleon exchange model for $\omega N$ TDAs.
COZ
\cite{Chernyak:1987nv}
(long-dashed lines) and
KS
\cite{King:1986wi}
(solid)
solutions for the leading twist nucleon DA
are used as the phenomenological input.}
\label{FigCSomegaDT2}
\end{figure}

On Fig.~\ref{FigCSomegaDT2} we show the
$\frac{d^2 \sigma_T}{d \Omega_\omega}$
cross section
as a function of
$\Delta_T^2$
for several values of $W$ and $Q^2$
({\it i.e.} for $u \le u_{0}$ or, equivalently, $\theta^*_\omega  \le \pi$).

We may conclude that qualitatively  the expected cross sections are similar to the backward $\pi$-electroproduction case.
Depending on the phenomenological input, and kinematics the cross section turns to be about $\sim 10 \div 100$ [nb/sr]
which lies within the reach of future JLab Hall A, B and C experiments.

\subsection*{Backward $\phi$ -meson hard photoproduction}
The second example is the  $\phi(1020)$ meson hard photoproduction
\be
\gamma^*(q,\lambda_\gamma) + p(p_1,s_1) \to p(p_2,s_2)+ \phi(p_\phi, \lambda_\phi).
\ee

Some controversy exists in the literature for the values of
the phenomenological $\phi$-meson to nucleon vector and tensor couplings.
As the numerical input for the $u$-channel nucleon exchange model for $\phi N$ TDAs
we employ the phenomenological $\phi$-meson to nucleon vector and tensor coupling presented in
Ref.~\cite{Mergell:1995bf} (see also Table~2 of Ref.~\cite{Meissner:1997qt}):
\be
G^V_{\phi NN}=9.18; \ \ \ G^T_{\phi NN}=-2.02,
\ee
that are roughly consistent with the estimates of Ref.~\cite{Hohler:1976ax}.

\begin{figure}[H]
 \begin{center}
 \epsfig{figure= 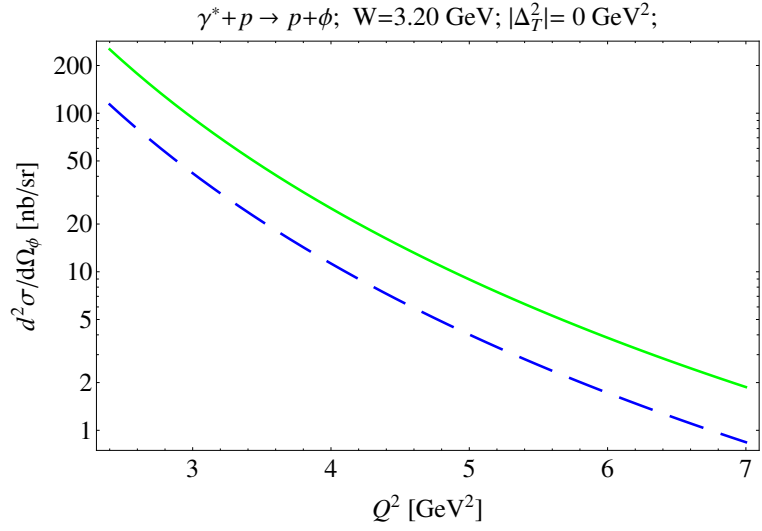 , height=7cm}
  \end{center}
  \caption{
Unpolarized cross section
$\frac{d^2 \sigma_T}{d \Omega_{\phi}}$
(in nb/sr) for backward
$\gamma^*+ p \to p+\phi$
for fixed $W=3.20$ GeV as
a  function of $Q^2$ in the $u$-channel nucleon exchange model for $\phi N$ TDAs.
COZ
\cite{Chernyak:1987nv}
(long-dashed lines) and
KS
\cite{King:1986wi}
(solid)
solutions for the leading twist nucleon DA
are used as the phenomenological input.
  }
\label{Fig_CSphiQ2}
\end{figure}

On Fig.~\ref{Fig_CSphiQ2} we plot the backward $\phi$-meson hard photoproduction cross section
$\frac{d^2 \sigma_T}{d \Omega_\phi}$
for
$\Delta_T^2=0$ ({\it i.e.} corresponding to exactly backward scattering $\theta^*_\phi=-\pi$)
as a function of $Q^2$ for the $Q^2$ range corresponding to
{\verb"Fpi-3 (E06-12-101)"}
$u$-channel kinematics
\cite{EMPworkshop}.

\begin{figure}[H]
 \begin{center}
  \epsfig{figure= 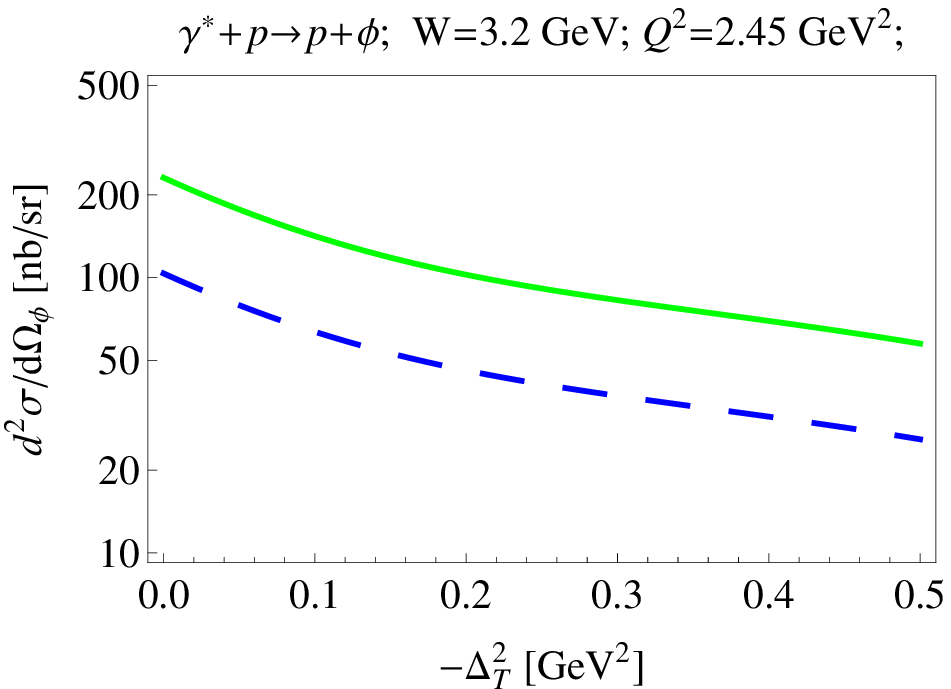 , height=5.5cm}
  \epsfig{figure= 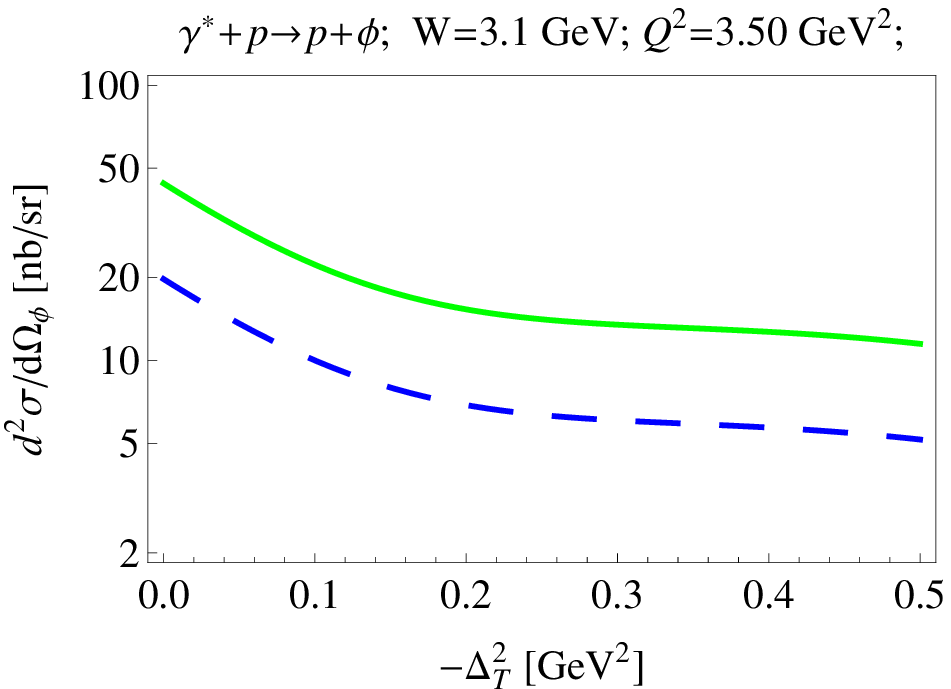 , height=5.5cm}
  \epsfig{figure= 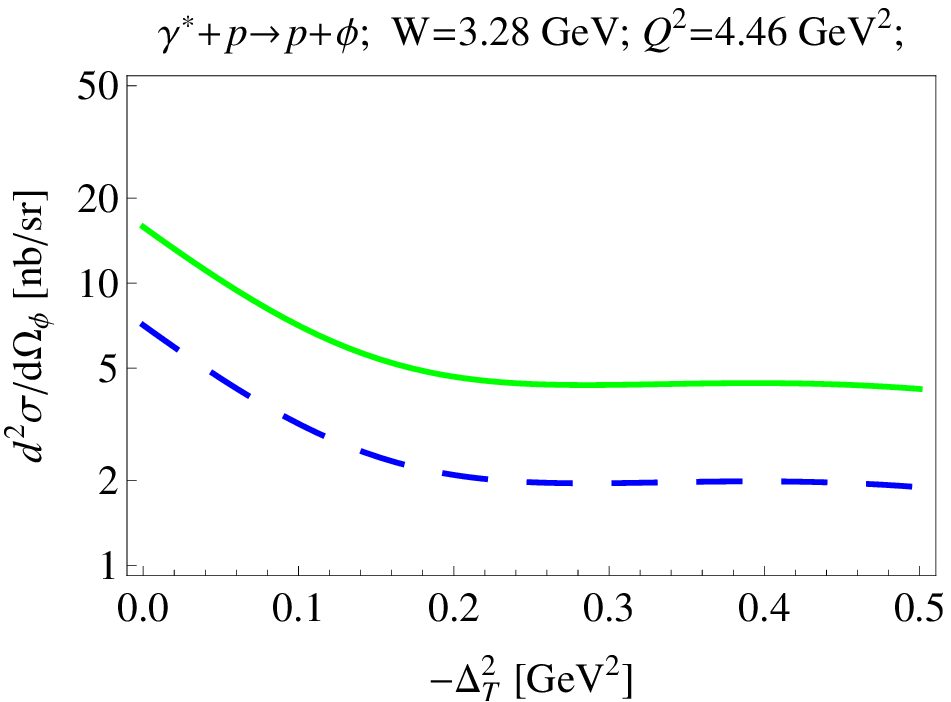 , height=5.5cm}
  \epsfig{figure= 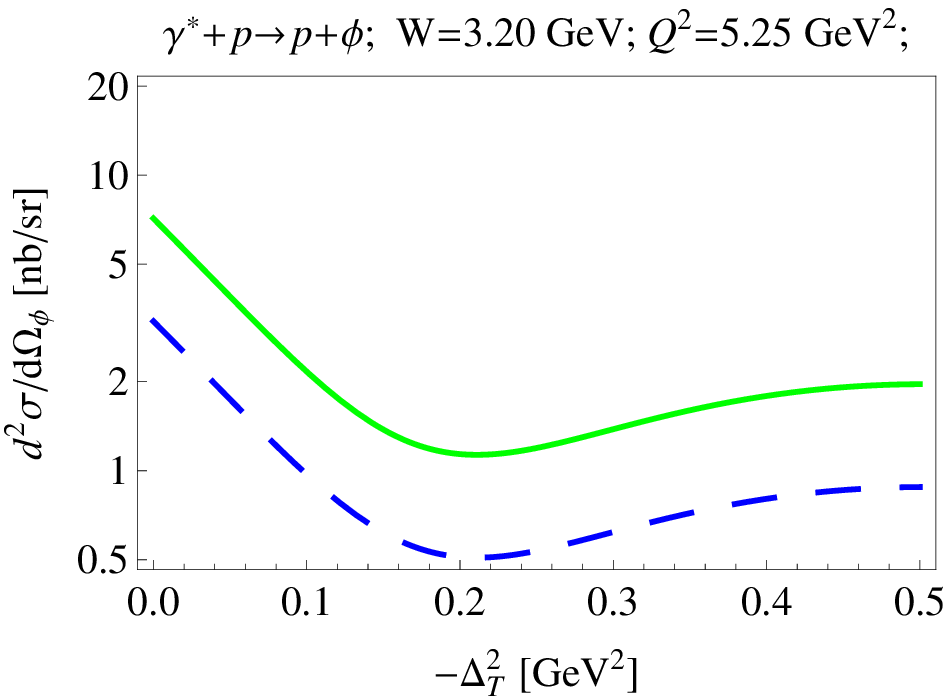 , height=5.5cm}
  \epsfig{figure= 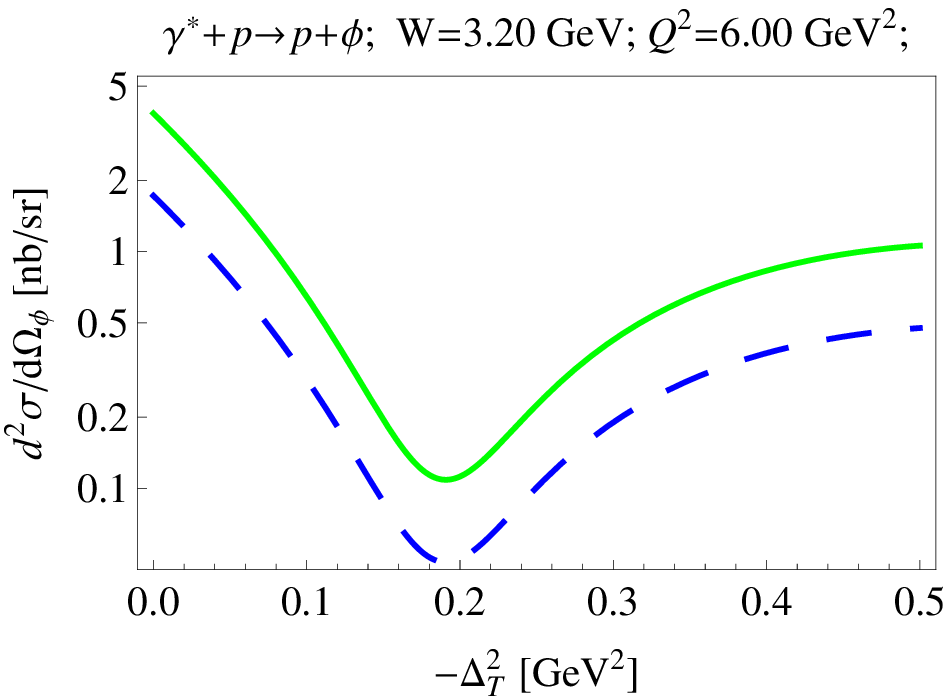 , height=5.5cm}
 \ \ \  \ \ \ \ \ \ \ \ \ \ \ \ \ \ \ \ \ \ \ \ \ \ \ \ \ \ \  \ \ \ \ \ \ \ \ \ \ \ \ \ \ \ \ \ \ \ \ \ \ \ \ \ \ \ \ \ \ \ \ \ \
 \end{center}
 \caption{Unpolarized cross section
$\frac{d^2 \sigma_T}{d \Omega_\phi}$
(in nb/sr) for backward
$\gamma^*  + p  \to p +\phi $
for several values of $W$ and $Q^2$
as a function of $\Delta_T^2$ in the $u$-channel nucleon exchange model for $\phi N$ TDAs.
COZ
\cite{Chernyak:1987nv}
(long-dashed lines) and
KS
\cite{King:1986wi}
(solid)
solutions for the leading twist nucleon DA
are used as the phenomenological input.}
 \label{Fig_CSphiDT2}
\end{figure}

On Fig.~\ref{Fig_CSphiDT2} we show the
$\frac{d^2 \sigma_T}{d \Omega_\phi}$
cross section
as a function of
$\Delta_T^2$
for several values of $W$ and $Q^2$
({\it i.e.} for $u \le u_{0}$ or, equivalently, $\theta^*_\phi  \le \pi$).

\subsection*{Backward $\rho^0$ -meson hard photoproduction}
Finally,
we consider the $\rho^0(770)$ meson hard photoproduction
\be
\gamma^*(q,\lambda_\gamma) + p(p_1,s_1) \to p(p_2,s_2)+\rho^0(p_\phi, \lambda_\phi).
\ee

As the numerical input for the $u$-channel nucleon exchange model for $\rho N$ TDAs
we employ the
{\verb"Pietarinen'77"}
phenomenological $\rho$-meson to nucleon vector and tensor couplings presented in
Table~9.2 of Ref.~\cite{Dumbrajs:1983jd}:
\be
G^V_{\rho NN}=2.6; \ \ \ G^T_{\rho NN}=16.1.
\label{rho_coupl}
\ee

\begin{figure}[H]
 \begin{center}
 \epsfig{figure= 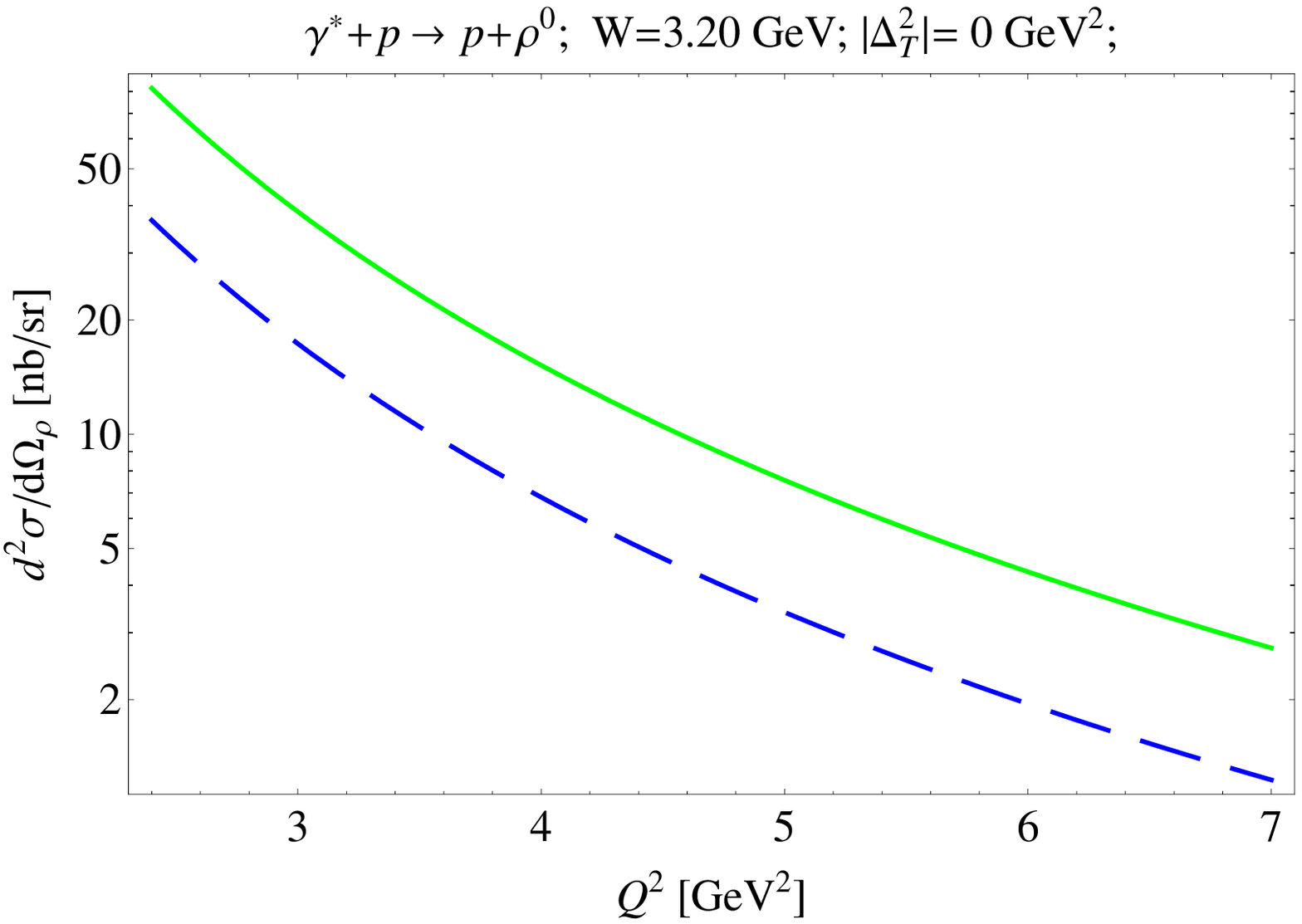 , height=7cm}
  \end{center}
  \caption{
Unpolarized cross section
$\frac{d^2 \sigma_T}{d \Omega_{\rho}}$
(in nb/sr) for backward
$\gamma^*  + p \to p +\rho^0$
for fixed $W=3.20$ GeV as
a  function of $Q^2$ in the $u$-channel nucleon exchange model for $\rho N$ TDAs.
COZ
\cite{Chernyak:1987nv}
(long-dashed lines) and
KS
\cite{King:1986wi}
(solid)
solutions for the leading twist nucleon DA
are used as the phenomenological input.
  }
\label{Fig_CSrhoQ2}
\end{figure}

On Fig.~\ref{Fig_CSrhoQ2} we plot the backward $\phi$-meson hard photoproduction cross section
$\frac{d^2 \sigma_T}{d \Omega_\rho}$
for
$\Delta_T^2=0$ ({\it i.e.} corresponding to exactly backward scattering $\theta^*_\rho=-\pi$)
as a function of $Q^2$ for the $Q^2$ range corresponding to
{\verb"Fpi-3 (E06-12-101)"}
$u$-channel kinematics
\cite{EMPworkshop}.

\begin{figure}[H]
 \begin{center}
  \epsfig{figure= 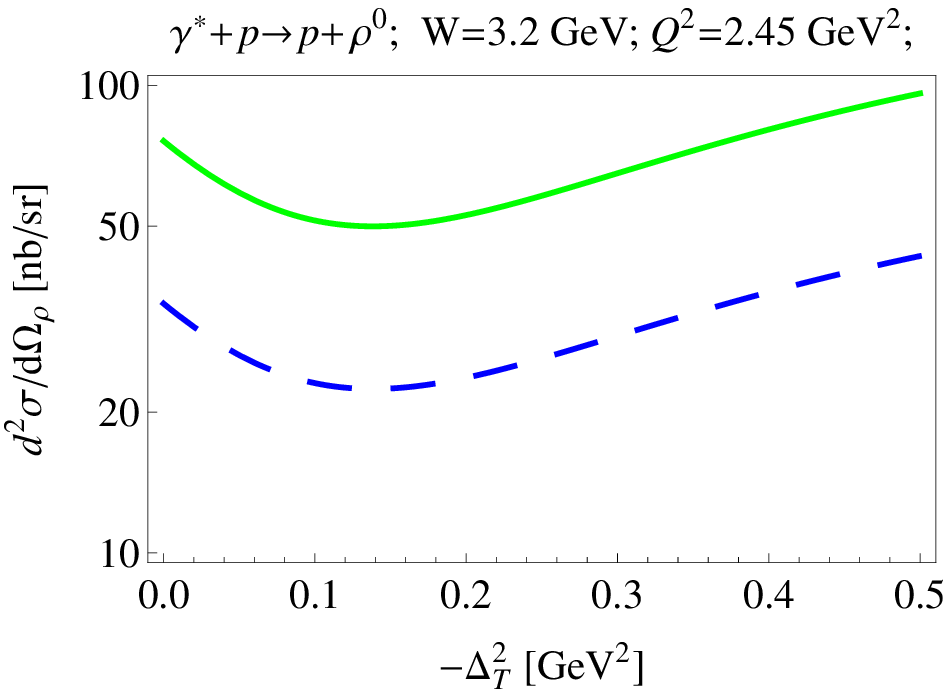 , height=5.5cm}
  \epsfig{figure= 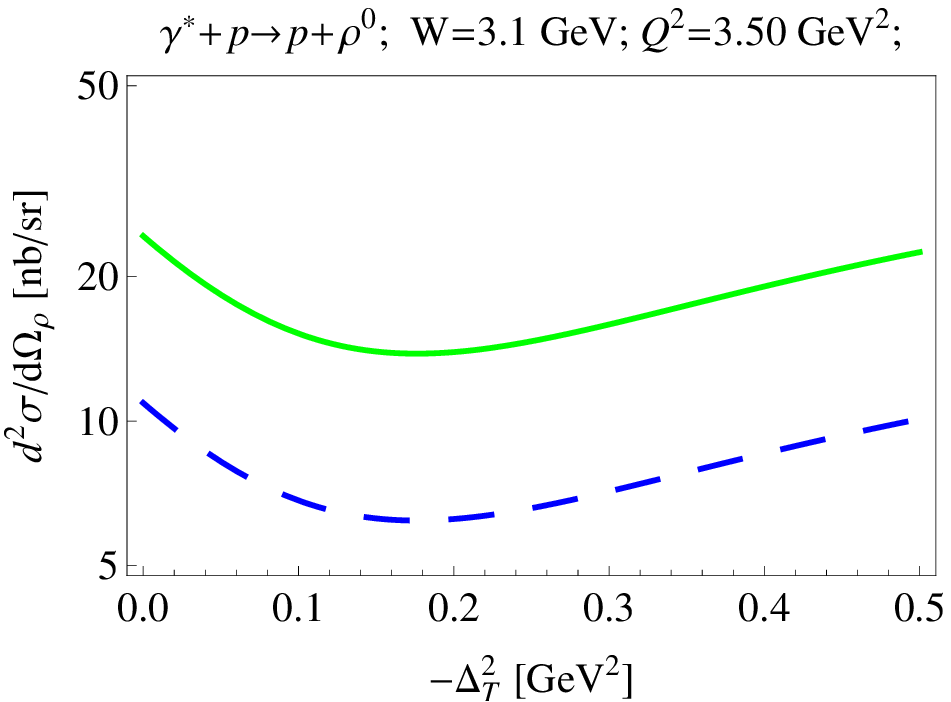 , height=5.5cm}
  \epsfig{figure= 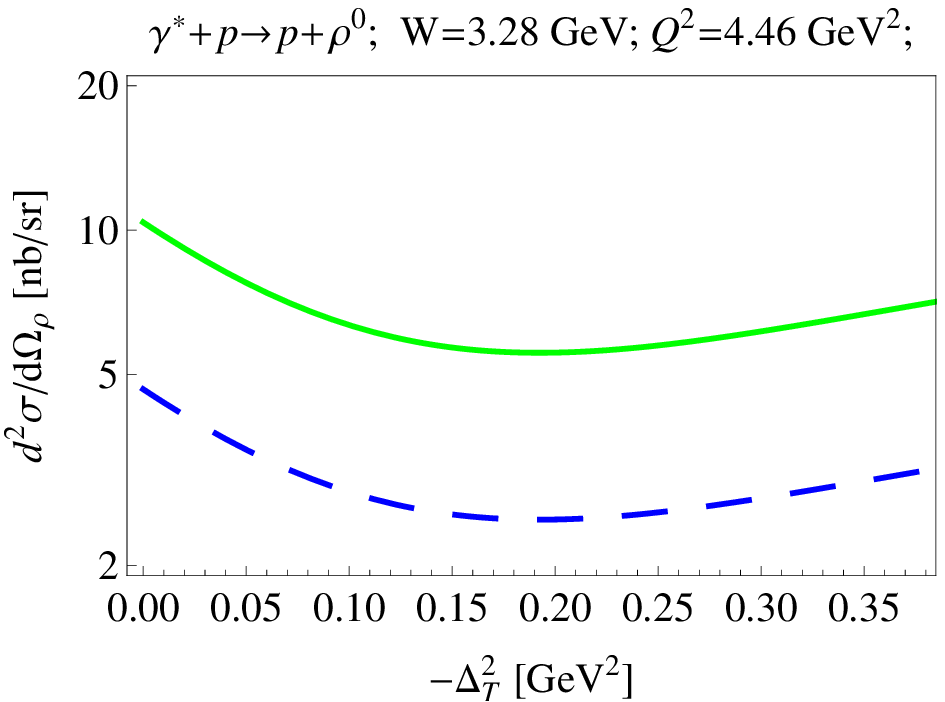 , height=5.5cm}
  \epsfig{figure= 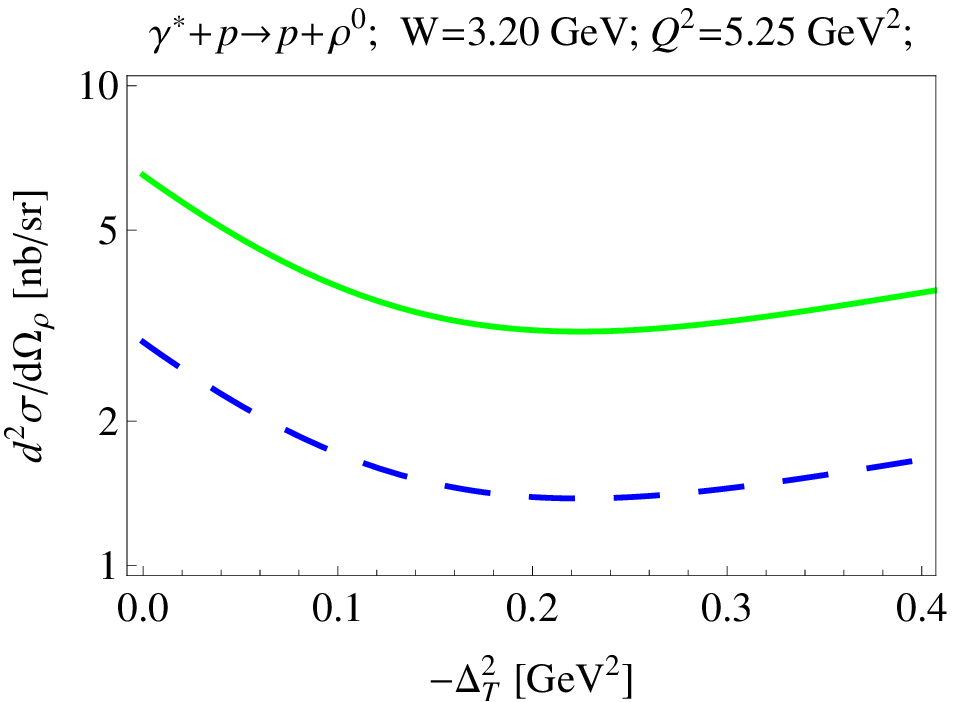 , height=5.5cm}
  \epsfig{figure= 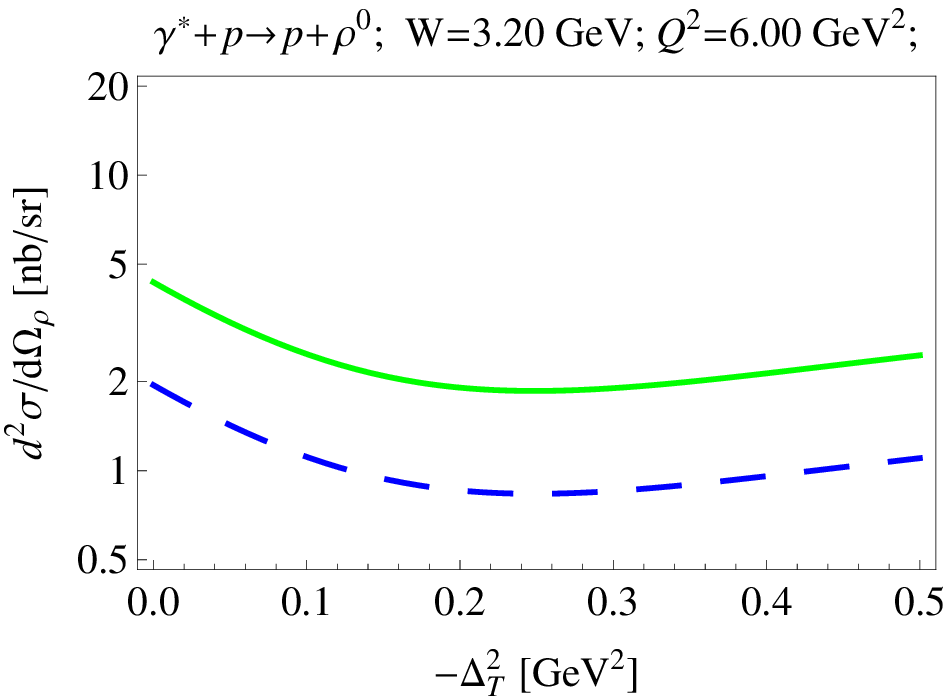 , height=5.5cm}
 \ \ \  \ \ \ \ \ \ \ \ \ \ \ \ \ \ \ \ \ \ \ \ \ \ \ \ \ \ \  \ \ \ \ \ \ \ \ \ \ \ \ \ \ \ \ \ \ \ \ \ \ \ \ \ \ \ \ \ \ \ \ \ \
 \end{center}
 \caption{Unpolarized cross section
$\frac{d^2 \sigma_T}{d \Omega_\rho}$
(in nb/sr) for backward
$\gamma^*  + p  \to p + \rho^0$
for several values of $W$ and $Q^2$
as a function of $\Delta_T^2$ in the $u$-channel nucleon exchange model for $\rho N$ TDAs. COZ
\cite{Chernyak:1987nv}
(long-dashed lines) and
KS
\cite{King:1986wi}
(solid)
solutions for the leading twist nucleon DA
are used as the phenomenological input.}
 \label{Fig_CSrhoDT2}
\end{figure}

On Fig.~\ref{Fig_CSrhoDT2} we show the
$\frac{d^2 \sigma_T}{d \Omega_\rho}$
cross section
as a function of
$\Delta_T^2$
for several values of $W$ and $Q^2$
({\it i.e.} for $u \le u_{0}$ or, equivalently, $\theta^*_\rho  \le \pi$).

It is interesting to note that as the $\rho$-meson is mostly coupled to nucleon through the tensor coupling
the cross section turns out to be suppressed at least by a factor $3$ as compared to the $\omega$ meson case.

\section{Conclusions and Outlook}
\label{Sec_Concl}

In this paper we applied the nucleon-to-meson TDA approach to the case of near-backward leptoproduction of
light vector mesons off nucleons. We defined $24$ leading twist-$3$ $VN$ TDAs and computed the corresponding
leading order hard amplitude. We estimated the differential cross section of backward $\rho^0$, $\omega$ and $\phi$
production within a simple cross-channel nucleon exchange model for $VN$ TDAs. The cross sections
were found to be sizable enough to be studied at future JLab experiments. The study of hard leptoproduction of
vector meson in the backward direction can also be seen as an opportunity for COMPASS at CERN and future EIC. Bringing experimental
evidences for the suggested scaling behavior will improve our understanding of the on-set of
perturbative QCD description of hard reactions.

The same $VN$ TDAs can also be addressed
in nucleon-antinucleon annihilation into a lepton pair in association with a vector meson to be studied at
\={P}ANDA, thus checking  the universality of TDAs. Therefore, a feasibility study for
$p \bar{p} \to \gamma^* V$ and $p \bar{p} \to J/\psi \, V$
reactions for the \={P}ANDA conditions is highly needed.

Our formalism can be naturally generalized to the case of nucleon-to-photon TDAs
\cite{Pire:2004ie},
which can be accessed in backward DVCS.
Potentially, this is the cleanest process involving  TDAs that could bring new information on hadronic structure.
However, the experimental feasibility of backward DVCS requires further studies.

\section*{Acknowledgements}
We would especially like to thank  Garth Huber for useful discussions on the preliminary
backward vector meson production data from  JLab Hall C.
We are also grateful to
Ermias Atomssa,
Michel Guidal,
Thierry Hennino,
Valery Kubarovsky,
Jean-Philippe Lansberg,
Frank Maas,
Maria Carmen Mora Espi,
Beatrice Ramstein,
Paul Stoler,
Christian Weiss
and
Manuel Zambrana
for valuable conversations on various aspects of baryon-to-meson TDA formalism and its experimental perspectives and to 
Vladimir Braun for useful correspondence. 
This work is partly supported by the
Polish Grant NCN No DEC-2011/01/B/ST2/03915, the
Joint Research Activity Study of Strongly Interacting Matter
(acronym HadronPhysics3, Grant 283286) under the Seventh Framework
Programme of the European Community and by the COPIN-IN2P3 Agreement and by
the French grant ANR PARTONS (ANR-12-MONU-0008-01).

\setcounter{section}{0}
\setcounter{equation}{0}
\renewcommand{\thesection}{\Alph{section}}
\renewcommand{\theequation}{\thesection\arabic{equation}}

\section{Isospin and permutation symmetry properties of  $VN$ TDAs}
\label{App_Isospin}

In this Appendix we present a brief overview of the isospin properties
of $VN$ TDAs. Throughout this analysis we literally follow the system of notations and conventions adopted in
Ref.~\cite{Pire:2011xv}.
\begin{itemize}
\item Letters from the beginning of the Greek alphabet are reserved for the
$SU(2)$ isospin indices
$ \alpha,\,\beta,\,\gamma,\, \iota, \, \kappa  ={1,\,2 }$.
\item We have to distinguish between upper (contravariant) and lower (covariant) $SU(2)$ isospin indices.
We introduce the totally antisymmetric tensor $\varepsilon_{\alpha \beta}$ for lowering indices and $\varepsilon^{\alpha \beta}$
for rising indices ($\varepsilon_{1 \,2}=\varepsilon^{1 \,2}=1$):
$\Psi ^\alpha \varepsilon_{\alpha \beta} = \Psi_\beta$,
$\Psi_\alpha \varepsilon^{\alpha \beta} = \Psi^\beta$
and
$\delta^\alpha_{\; \beta}= -\varepsilon^\alpha_{\; \beta}= \varepsilon_\beta^{\; \, \alpha}$.
\item Letters from the beginning of the Latin alphabet
$a,b,c=1,\,2,\,3$
are reserved for indices of the adjoint representation of the $SU(2)$ isospin group. $\sigma_a$ stand for the usual
Pauli matrices.
\item Letters from the second half of the Greek alphabet $\rho, \, \tau,\, \chi$ are reserved for the Dirac indices.
\item Letters $c_1$, $c_2$, $c_3$ stand for $SU(3)$ color indices.
\end{itemize}
We consider the $VN$ matrix element of light-cone three-quark operators
\be
\hat{O}_{\rho \tau \chi}^{\alpha \beta \gamma}(z_1,z_2,z_3)
\equiv
\hat{O}_{\rho \tau \chi}^{\alpha \beta \gamma}(1,2,3)
= \varepsilon_{c_1 c_2 c_3} \Psi_\rho^{c_1 \alpha}(z_1) \Psi_\tau^{c_2 \beta}(z_2) \Psi_\chi^{c_3 \alpha}(z_3).
\ee

For the case of $I=0$ vector meson ($\omega(782)$ and $\phi(1020)$ being the examples) the isotopic structure of $VN$ TDA
coincides with that of the leading twist nucleon DA. Therefore, the invariant isospin parametrization reads
(for definiteness we consider the $\omega N$ TDA case)
\be
4 \langle \omega(p_\omega, \lambda_\omega)| \hat{O}_{\rho \tau \chi}^{\alpha \beta \gamma}(1,2,3)| N_{\iota}(p_1, s_1) \rangle=
\varepsilon^{\alpha \beta} \delta^\gamma_\iota M^{\omega N \{12 \}}_{\rho \chi \tau}(1,3,2) +
\varepsilon^{\alpha \gamma} \delta^\gamma_\iota M^{\omega N \{12 \}}_{\rho \tau \chi}(1,2,3),
\label{Isospin_Par_I=0}
\ee
where the invariant amplitude $ M^{\omega N \{12 \}}_{\rho \tau \chi}(1,2,3)$ is symmetric with respect to interchange of first two variables:
\be
M^{\omega N \{12 \}}_{\rho \tau \chi}(1,2,3)=M^{\omega N \{12 \}}_{\tau \rho \chi}(2,1,3),
\ee
and satisfies the isospin identity ({\it c.f.} eq. (32) of Ref.~\cite{Pire:2011xv})
\be
M^{\omega N \{12 \}}_{\rho \tau \chi}(1,2,3)+ M^{\omega N \{12 \}}_{\rho \chi \tau}(1,3,2)+M^{\omega N \{12 \}}_{\tau \chi \rho}(2,3,1)=0.
\label{Isospin_Id_omegaN}
\ee
The Dirac structure of
$M^{\omega N\,\{12\}}_{\rho \tau \chi}(1,2,3)$
is that of
eq.~(\ref{VN_TDAs_param}). It is straightforward to check that
\be
&&
4 \langle \omega(p_\omega, \lambda_\omega)| \hat{O}_{\rho \tau \chi}^{uud}(1,2,3)| N_{p}(p_1, s_1) \rangle=
-4 \langle \omega(p_\omega, \lambda_\omega)| \hat{O}_{\rho \tau \chi}^{ddu}(1,2,3)| N_{n}(p_1, s_1) \rangle
\nonumber \\ &&
=M^{\omega N\,\{12\}}_{\rho \tau \chi}(1,2,3).
\ee
To work out the consequences of the isospin identity
(\ref{Isospin_Id_omegaN})
for particular TDAs of
(\ref{VN_TDAs_param})
one has to employ the set of the Fierz identities
(\ref{Fiers_1E_set})--(\ref{Fiers_last_set})
for the relevant Dirac structures
(\ref{Dirac_v_NV}),
(\ref{Dirac_a_NV}), (\ref{Dirac_t_NV}).

For the case of $I=1$ vector meson ($\rho(770)$ being the obvious example) the isotopic structure of $VN$ TDA
coincides with that of the leading twist nucleon $\pi N$ TDA.
Therefore we can write down the following isospin decomposition
\be
&&
4\langle  \rho_a | \widehat{O}^{\alpha \beta \gamma}_{\rho \tau \chi}(z_1,\,z_2,\,z_3) | N_\iota \rangle
=(f_a)^{\{\alpha \beta \gamma\}}_{\ \ \ \ \iota} M^{(\rho N)_{3/2}}_{\rho \tau \chi}
(1,2,3)
\nonumber \\ && +
\varepsilon^{\alpha \beta} (\sigma_a)^\gamma_{\ \iota}  M^{(\rho N)_{1/2} \, \{1 2 \}}_{\rho \chi \tau}
(1,3,2)
+\varepsilon^{\alpha \gamma} (\sigma_a)^\beta_{\ \iota} M^{(\rho N)_{1/2} \, \{1 2 \}}_{\rho  \tau \chi}
(1,2,3)\,,
\label{rhoN_isospin_dec}
\ee
where the totally symmetric tensor
$(f_a)^{\{\alpha \beta \gamma\}}_{\ \ \ \ \iota}$
reads
\be
&&
(f_a)^{\{\alpha \beta \gamma\}}_{\ \ \ \ \iota}
= \frac{1}{3}
\left(
(\sigma_a)^{\alpha}_{\; \delta} \varepsilon^{\delta \beta} \delta^{\gamma}_{\; \iota}
+
(\sigma_a)^{\alpha}_{\; \delta} \varepsilon^{\delta \gamma} \delta^{\beta}_{\; \iota}+
(\sigma_a)^{\beta}_{\; \delta} \varepsilon^{\delta \gamma} \delta^{\alpha}_{\; \iota}
\right).
\label{Def_ta_tensor}
\ee

The properties of the $u$-channel isospin-$\frac{1}{2}$ invariant amplitude
$M^{(\rho N)_{1/2} \, \{1 2 \}}_{\rho  \tau \chi}
$
are fully analogous to that of the corresponding invariant amplitude of
eq.~(\ref{Isospin_Par_I=0}).
It is symmetric under the simultaneous interchange of two first arguments and the Dirac indices:
\be
M^{(\rho N)_{1/2} \, \{1 2 \}}_{\rho \tau \chi}(1,2,3)=M^{(\rho N)_{1/2} \, \{1 2 \}}_{\tau \rho \chi}(2,1,3),
\ee
and satisfies the isospin identity ({\it c.f.} eq. (32) of Ref.~\cite{Pire:2011xv})
\be
M^{(\rho N)_{1/2} \, \{1 2 \}}_{\rho \tau \chi}(1,2,3)+ M^{(\rho N)_{1/2} \, \{1 2 \}}_{\rho \chi \tau}(1,3,2)+M^{(\rho N)_{1/2} \, \{1 2 \}}_{\tau \chi \rho}(2,3,1)=0.
\label{Isospin_Id_rhoN}
\ee
The Dirac structure of
$M^{(\rho N)_{1/2} \, \{1 2 \}}_{\rho \tau \chi}
$
is again that of
eq.~(\ref{VN_TDAs_param}).

As a consequence of the permutation and isotopic symmetry, the $u$-channel isospin-$\frac{3}{2}$ invariant amplitude
$M^{(\rho N)_{3/2} \, \{1 2 \}}_{\rho  \tau \chi}$
is   completely symmetric  under simultaneous permutations of
the arguments and the Dirac indices:
\be
&&
M^{(\rho N)_{3/2}}_{\rho \tau \chi} (1,2,3)
=
M^{(\rho N)_{3/2}}_{ \rho \chi \tau} (1,3,2)
=
M^{(\rho N)_{3/2}}_{\tau \rho \chi} (2,1,3)
 \nonumber \\ &&
=
M^{(\rho N)_{3/2}}_{ \tau \chi \rho} (2,3,1)
=
M^{(\rho N)_{3/2}}_{  \chi \tau \rho} (3,2,1)
=
M^{(\rho N)_{3/2}}_{ \chi \rho  \tau} (3,1,2)
 \,.
 \label{symmetries_MI32rhoNTDA}
\ee
The Dirac structure of
$M^{(\rho N)_{3/2} \, \{1 2 \}}_{\rho \tau \chi}
$
is also that of
eq.~(\ref{VN_TDAs_param}).

Below, to the leading twist-$3$ accuracy, we present the set of  Fierz identities for the relevant Dirac structures
(\ref{Dirac_v_NV}),
(\ref{Dirac_a_NV}), (\ref{Dirac_t_NV})
needed to establish the consequences of the isotopic and permutation symmetry
(\ref{symmetries_MI32rhoNTDA})
for $VN$ TDAs.
\be
&&
(v_{1 {\cal E}}^{V N})_{\rho \tau, \, \chi}= \frac{1}{2} \left(
 v_{1 {\cal E}}^{VN}-  a_{1 {\cal E}}^{VN}+
  t_{1 {\cal E}}^{VN}+  t_{2 {\cal E}}^{VN} \right)_{\chi \tau, \, \rho};
\nonumber \\ &&
(a_{1 {\cal E}}^{VN})_{\rho \tau, \, \chi}=\frac{1}{2} \left(
- v_{1 {\cal E}}^{VN} + a_{1 {\cal E}}^{VN} +
   t_{1 {\cal E}}^{VN} + t_{2 {\cal E}}^{VN}  \right)_{\chi \tau, \, \rho};
\nonumber \\ &&
(t_{1 {\cal E}}^{VN})_{\rho \tau, \, \chi}=
\frac{1}{2} \left(   v_{1 {\cal E}}^{VN} +
 a_{1 {\cal E}}^{VN} +
 t_{1 {\cal E}}^{VN} -
 t_{2 {\cal E}}^{VN}  \right)_{\chi \tau, \, \rho}
\nonumber \\ &&
(t_{2 {\cal E}}^{VN})_{\rho \tau, \, \chi}=\frac{1}{2} \left( v_{1 {\cal E}}^{VN} +
 a_{1 {\cal E}}^{VN} -
 t_{1 {\cal E}}^{VN} +
   t_{2 {\cal E}}^{VN} \right)_{\chi \tau, \, \rho};
\label{Fiers_1E_set}
\ee
\be
 &&
(v_{1T}^{VN})_{\rho \tau,\,\chi}= \frac{1}{2} \left(v_{1T}^{VN} - a_{1T}^{VN} + t_{1T}^{VN} \right)_{\chi \tau,\,\rho};
\nonumber \\ &&
(a_{1T}^{VN})_{\rho \tau,\,\chi}=  \frac{1}{2} \left(- v_{1T}^{VN}+ a_{1T}^{VN} +   t_{1T}^{VN} \right)_{\chi \tau,\,\rho};
\nonumber \\ &&
(t_{1T}^{VN})_{\rho \tau,\,\chi}=   \left(v_{1T}^{VN} +  (a_{1T}^{VN} \right)_{\chi \tau,\,\rho};
\label{Fiers_1T_set}
\ee
\be
&&
{(v_{1n}^{VN})}_{\rho \tau, \, \chi}= \frac{1}{2} \left(v_{1n}^{VN}-  a_{1n}^{VN} +  t_{1n}^{VN} \right)_{\chi \tau, \, \rho};
\; \nonumber \\ &&
{(a_{1n})}_{\rho \tau, \, \chi}= \frac{1}{2} \left(-v_{1n}^{VN}+  a_{1n}^{VN} +  t_{1n}^{VN} \right)_{\chi \tau, \, \rho};
\nonumber \\ &&
{(t_{1n}^{VN})}_{\rho \tau, \, \chi}=   \left( v_{1n}^{VN}+  a_{1n}^{VN} \right)_{\chi \tau, \, \rho};
\label{Fiers_1n_set}
\ee
\be
&&
(v^{VN}_{2 {\cal E}})_{\rho \tau,\,\chi}= \frac{1}{2}  \left( v^{VN}_{2 {\cal E}} -  a^{VN}_{2 {\cal E}} +
 t^{VN}_{3 {\cal E}}  -t^{VN}_{4 {\cal E}} \right)_{\chi \tau,\,\rho}; \nonumber \\ &&
(a^{VN}_{2 {\cal E}})_{\rho \tau,\,\chi}=  \frac{1}{2} \left(-v^{VN}_{2 {\cal E}} +  a^{VN}_{2 {\cal E}}
+ t^{VN}_{3 {\cal E}} - t^{VN}_{4 {\cal E}} \right)_{\chi \tau,\,\rho}; \nonumber \\ &&
(t^{VN}_{3 {\cal E} })_{\rho \tau,\,\chi}= \frac{1}{2 }  \left(v_{1T}^{VN}  +     v_{2 {\cal E}}^{VN} +
  a_{1T}^{VN}  +     a_{2\varepsilon}^{VN}  +
 t_{3{\cal E}}^{VN} -  t_{1T}^{VN} +  t_{4 {\cal E}}^{VN} \right)_{\chi \tau,\,\rho};
\nonumber \\ &&
(t^{VN}_{4{\cal E}})_{\rho \tau,\,\chi}= \frac{1}{2} \left( - v^{VN}_{2 {\cal E}}+  v^{VN}_{1 T}
-  a^{VN}_{2 {\cal E}}+ a^{VN}_{1 T}
+   t^{VN}_{3{\cal E}}-  t^{VN}_{1T} +  t^{VN}_{4{\cal E}} \right)_{\chi \tau,\,\rho};
\label{Fiers_2E_set}
\ee
\be
&&
(v_{2T}^{VN})_{\rho \tau,\,\chi}=\frac{1}{2} \left(v_{2T}^{VN} -   a_{2T}^{VN} + t_{2T}^{VN} + t_{3T}^{VN}
\right)_{\chi \tau,\,\rho};
\nonumber \\ &&
(a_{2T}^{VN})_{\rho \tau,\,\chi}=\frac{1}{2} \left(-v_{2T}^{VN} + a_{2T}^{VN} + t_{2T}^{VN} + t_{3T}^{VN}
\right)_{\chi \tau,\,\rho};
\nonumber \\ &&
(t_{2T}^{VN})_{\rho \tau,\,\chi}=\frac{1}{2} \left(
v_{2T}^{VN} + a_{2T}^{VN} + t_{2T}^{VN} - t_{3T}^{VN}
\right)_{\chi \tau,\,\rho};
\nonumber \\ &&
(t_{3T}^{VN})_{\rho \tau,\,\chi}=\frac{1}{2} \left(
v_{2T}^{VN} + a_{2T}^{VN} - t_{2T}^{VN} + t_{3T}^{VN}
\right)_{\chi \tau,\,\rho};
\label{Fiers_2T_set}
\ee
\be
&&
(v_{2n}^{VN})_{\rho \tau,\,\chi}=\frac{1}{2} \left(v_{2n}^{VN} -   a_{2n}^{VN} + t_{2n}^{VN} + t_{3n}^{VN}
\right)_{\chi \tau,\,\rho};
\nonumber \\ &&
(a_{2n}^{VN})_{\rho \tau,\,\chi}=\frac{1}{2} \left(-v_{2n}^{VN} + a_{2n}^{VN} + t_{2n}^{VN} + t_{3n}^{VN}
\right)_{\chi \tau,\,\rho};
\nonumber \\ &&
(t_{2n}^{VN})_{\rho \tau,\,\chi}=\frac{1}{2} \left(
v_{2n}^{VN} + a_{2n}^{VN} + t_{2n}^{VN} - t_{3n}^{VN}
\right)_{\chi \tau,\,\rho};
\nonumber \\ &&
(t_{3n}^{VN})_{\rho \tau,\,\chi}=\frac{1}{2} \left(
v_{2n}^{VN} + a_{2n}^{VN} - t_{2n}^{VN} + t_{3n}^{VN}
\right)_{\chi \tau,\,\rho};
\label{Fiers_2n_set}
\ee
\be
&&
(t_{4T}^{VN})_{\rho \tau,\,\chi}= \frac{1}{2}  \frac{\Delta_T^2}{M^2} \left(v_{1T}^{VN}  -   a_{1T}^{VN}
-   t_{1T}^{VN} \right)_{\chi \tau,\,\rho}+(t_{4T}^{VN})_{\chi \tau,\,\rho}; \nonumber \\ &&
(t_{4n}^{VN})_{\rho \tau,\,\chi}= \frac{1}{2}  \frac{\Delta_T^2}{M^2} \left(v_{1n}^{VN}  -   a_{1n}^{VN}
-   t_{1n}^{VN} \right)_{\chi \tau,\,\rho}+(t_{4n}^{VN})_{\chi \tau,\,\rho}.
 \label{Fiers_last_set}
\ee

\setcounter{equation}{0}

\section{Nucleon pole exchange model for $V N$ TDA}
\label{App_NPole}
In this Appendix we construct a simple $u$-channel nucleon
exchange model for $VN$ TDAs. This model populates only
the Efremeov-Radyushkin-Brodsky-Lepage (ERBL)-like region
of the $VN$ TDA support domain and represents an analogue of
the $D$-term contribution supplementary to the spectral representation
\cite{Pire:2010if}
in terms of quadruple distributions.

The effective $V N N$ vertex
\cite{Dumbrajs:1983jd}
\be
&&
V_{\it eff} \big( N(p_1,s_1) \to V(p_V, \lambda_V) N(-\Delta,s_p) \big)
\nonumber \\ &&
=\bar{U}(-\Delta, s_p) \left[ G^V_{VNN} \hat{{\cal E}}^*(p_V, \lambda_V)+ G^T_{VNN} \frac{\sigma_{\mu \nu}}{2M}
(-\Delta)^\nu  {{\cal E}^\mu}^*(p_V, \lambda_V)
\right] U(p_1, s_1)
\label{GTGV_couplings}
\ee
involves two dimensionless phenomenological couplings $G^V_{VNN}$ and $G^T_{VNN}$.

The $u$-channel nucleon exchange contribution into the light-cone three-quark-operator $VN$ matrix element
occurring in the $VN$ TDA definition (\ref{VN_TDAs_param}) reads:
\be
&&
\left.\langle V(p_V, \lambda_V)|
\hat{O}_{\rho \tau \chi}^{uud}(\lambda_1n, \, \lambda_2n, \, \lambda_3n)| N(p_1,s_1) \rangle \right|_{N(940)}
\nonumber \\ &&
=\sum_{s_p} \langle 0
|
\hat{O}_{\rho \tau \chi}^{uud}(\lambda_1n, \, \lambda_2n, \, \lambda_3n)| N(-\Delta,s_p) \rangle
\bar{U}(-\Delta, s_p)
\frac{1}{\Delta^2-M^2}
\left[ G^V_{V NN } \hat{ \cal E}^*(p_V, \lambda_V)
\right. \nonumber \\ &&
\left. + G^T_{NN\omega} \frac{\sigma_{\mu \nu}}{2M}
(-\Delta)^\nu  {{\cal E}^\mu}^*(p_V, \lambda_V)
\right] U(p_1, s_1).
\ee
The
$\langle 0|\hat{O}_{\rho \tau \chi}^{uud}(\lambda_1n, \, \lambda_2n, \, \lambda_3n)| N(-\Delta,s_p) \rangle$
matrix element is then expressed through the leading twist-$3$ nucleon DA.
Performing the Fourier transform (\ref{Fourier_TDA}) this yields:
\be
&&
4 {\cal F}(x_1,x_2,x_3) \left.\langle V(p_V, \lambda_V)|
\hat{O}_{\rho \tau \chi}^{uud}(\lambda_1n, \, \lambda_2n, \, \lambda_3n)| N_p(p_1,s_1) \rangle \right|_{N(940)}
= \delta(x_1+x_2+x_3-2\xi) \nonumber \\ && \times M  \left.
\Theta_{\rm ERBL}(x_1,x_2,x_3)
f_N \frac{1}{M} \right.   \times  \frac{1}{(2 \xi)^2} \sum_{s_p}
\Big\{
V^p \left(\frac{x_1}{2 \xi}, \frac{x_2}{2 \xi},\frac{x_3}{2 \xi} \right) (-\hat{\Delta} C)_{\rho \tau} (\gamma^5 U(-\Delta,s_p))_\chi
\nonumber \\ &&
+
A^p \left(\frac{x_1}{2 \xi}, \frac{x_2}{2 \xi},\frac{x_3}{2 \xi} \right) (-\hat{\Delta} \gamma_5 C)_{\rho \tau} U(-\Delta,s_p)_\chi
\nonumber \\ &&
+
T^p \left(\frac{x_1}{2 \xi}, \frac{x_2}{2 \xi},\frac{x_3}{2 \xi} \right) (\sigma_{-\Delta \lambda}  C)_{\rho \tau} ( \gamma^\lambda \gamma^5 U(-\Delta,s_p))_\chi
\Big\}
\bar{U}(-\Delta, s_p)
\frac{1}{\Delta^2-M^2}
\left[ G^V_{V NN } \hat{\cal E}^*(p_V, \lambda_V)
\right. \nonumber \\ &&
\left. + G^T_{V NN } \frac{\sigma_{\mu \nu}}{2M}
(-\Delta)^\nu  {{\cal E}^\mu}^*(p_V, \lambda_V)
\right] U(p_1, s_1).
\label{Nucleon_exchange_VNTDA}
\ee
Here we employ the notation
\be
\Theta_{\rm ERBL}(x_1,x_2,x_3) \equiv \left[ \prod_{k=1}^3 \theta(0 \le x_k \le 2\xi) \right].
\ee
To work out from
(\ref{Nucleon_exchange_VNTDA})
the nucleon pole exchange contributions  to particular TDAs
one has to expand it over the set of
$24$
basic Dirac structures
(\ref{Dirac_v_NV}),
(\ref{Dirac_a_NV})
and
(\ref{Dirac_t_NV})
which is a straightforward (though tedious) calculation.
It turns out that the $u$-channel nucleon exchange model populates
$22$
out of the
$24$
$VN$
TDAs
($T_{4T}^{VN}$ and $T_{4n}^{VN}$ vanish in this model).

It is convenient to show the results for the groups of $VN$ TDAs interlinked through the
set of the isospin relations (see Appendix~\ref{App_Isospin}).
\bi
\item $V_{1 {\cal E}}$, $A_{1 {\cal E}}$, $T_{1 {\cal E}}$, $T_{2 {\cal E}}$
satisfy the isospin symmetry relations based on the Fierz transformation set (\ref{Fiers_1E_set}).
\be
&&
\left. V_{1 {\cal E}}^{VN}(x_1, x_2, x_3, \xi, \Delta^2) \right|_{N(940)} = \Theta_{\rm ERBL}(x_1,x_2,x_3) \frac{1}{(2 \xi)^2} V^p \left(  \frac{x_1}{2 \xi}, \frac{x_2}{2 \xi}, \frac{x_3}{2 \xi}  \right)
K_{1 {\cal E}}^{VN}( \xi, \Delta^2); \nonumber \\ &&
\left. A_{1 {\cal E}}^{VN}(x_1, x_2, x_3, \xi, \Delta^2) \right|_{N(940)} = \Theta_{\rm ERBL}(x_1,x_2,x_3) \frac{1}{(2 \xi)^2} A^p \left(  \frac{x_1}{2 \xi}, \frac{x_2}{2 \xi}, \frac{x_3}{2 \xi}  \right)
K_{1 {\cal E}}^{VN}( \xi, \Delta^2); \nonumber  \\ &&
\left. T_{1 {\cal E}}^{VN}(x_1, x_2, x_3, \xi, \Delta^2)\right|_{N(940)}= -\Theta_{\rm ERBL}(x_1,x_2,x_3) \frac{1}{(2 \xi)^2} T^p \left(  \frac{x_1}{2 \xi}, \frac{x_2}{2 \xi}, \frac{x_3}{2 \xi}  \right)
K_{1 {\cal E}}^{VN}( \xi, \Delta^2); \nonumber  \\ &&
\left. T_{2 {\cal E}}^{VN}(x_1, x_2, x_3, \xi, \Delta^2)\right|_{N(940)}= -\Theta_{\rm ERBL}(x_1,x_2,x_3) \frac{1}{(2 \xi)^2} T^p \left(  \frac{x_1}{2 \xi}, \frac{x_2}{2 \xi}, \frac{x_3}{2 \xi}  \right)
K_{1 {\cal E}}^{VN}( \xi, \Delta^2), \nonumber  \\ &&
\ee
where
\be
K_{1 {\cal E}}^{VN}( \xi, \Delta^2)= \frac{f_N}{\Delta^2-M^2} \left( G^V_{VNN} \frac{2\xi(1-\xi)}{1+\xi} + G^T_{VNN}\,  \xi  \left(\frac{2 \xi }{1+\xi }-\frac{ \Delta^2}{M^2}\right) \right).
\ee

\item $V_{1 T}$, $A_{1 T}$, $T_{1 T}$ satisfy the isospin symmetry relations based on the Fierz transformation set (\ref{Fiers_1T_set})
\be
&&
\left. V_{1 T}^{VN}(x_1, x_2, x_3, \xi, \Delta^2) \right|_{N(940)} = \Theta_{\rm ERBL}(x_1,x_2,x_3) \frac{1}{(2 \xi)^2} V^p \left(  \frac{x_1}{2 \xi}, \frac{x_2}{2 \xi}, \frac{x_3}{2 \xi}  \right)
K_{1 T}^{VN}( \xi, \Delta^2); \nonumber \\ &&
\left. A_{1 T}^{VN}(x_1, x_2, x_3, \xi, \Delta^2) \right|_{N(940)} = \Theta_{\rm ERBL}(x_1,x_2,x_3) \frac{1}{(2 \xi)^2} A^p \left(  \frac{x_1}{2 \xi}, \frac{x_2}{2 \xi}, \frac{x_3}{2 \xi}  \right)
K_{1 T}^{VN}( \xi, \Delta^2); \nonumber  \\ &&
\left. T_{1 T}^{VN}(x_1, x_2, x_3, \xi, \Delta^2)\right|_{N(940)}= -\Theta_{\rm ERBL}(x_1,x_2,x_3) \frac{1}{(2 \xi)^2} T^p \left(  \frac{x_1}{2 \xi}, \frac{x_2}{2 \xi}, \frac{x_3}{2 \xi}  \right)
K_{1 T}^{VN}( \xi, \Delta^2),
\nonumber  \\ &&
\ee
where
\be
K_{1 T}( \xi, \Delta^2)= \frac{f_N}{\Delta^2-M^2} \left( -G^V_{VNN} \frac{2 \xi  (1+3 \xi)}{1-\xi}   \right).
\ee

\item $V_{1 n}$, $A_{1 n}$, $T_{1 n}$ satisfy the isospin symmetry relations based on the Fierz transformation set (\ref{Fiers_1n_set})
\be
&&
\left. V_{1 n}^{VN}(x_1, x_2, x_3, \xi, \Delta^2) \right|_{N(940)} = \Theta_{\rm ERBL}(x_1,x_2,x_3) \frac{1}{(2 \xi)^2} V^p \left(  \frac{x_1}{2 \xi}, \frac{x_2}{2 \xi}, \frac{x_3}{2 \xi}  \right)
K_{1 n}^{VN}( \xi, \Delta^2); \nonumber \\ &&
\left. A_{1 n}^{VN}(x_1, x_2, x_3, \xi, \Delta^2) \right|_{N(940)} = \Theta_{\rm ERBL}(x_1,x_2,x_3) \frac{1}{(2 \xi)^2} A^p \left(  \frac{x_1}{2 \xi}, \frac{x_2}{2 \xi}, \frac{x_3}{2 \xi}  \right)
K_{1 n}^{VN}( \xi, \Delta^2); \nonumber  \\ &&
\left. T_{1 n}^{VN}(x_1, x_2, x_3, \xi, \Delta^2)\right|_{N(940)}= -\Theta_{\rm ERBL}(x_1,x_2,x_3) \frac{1}{(2 \xi)^2} T^p \left(  \frac{x_1}{2 \xi}, \frac{x_2}{2 \xi}, \frac{x_3}{2 \xi}  \right)
K_{1 n}^{VN}( \xi, \Delta^2),
\nonumber  \\ &&
\ee
where
\be
&&
K_{1 n}( \xi, \Delta^2)
\nonumber \\ &&
= \frac{f_N}{\Delta^2-M^2}
\left(
\frac{(1+\xi) \left(m^2-\Delta_T^2\right)}{M^2 (1-\xi)^2}-\frac{1}{1+\xi} \right)
\left( G^V_{VNN}(-4 \xi) + G^T_{VNN} \frac{\xi (1-\xi)}{1+\xi}
\right).
\nonumber \\ &&
\ee

\item $V_{2 {\cal E}}$, $A_{2 {\cal E}}$, $T_{3 {\cal E}}$, $T_{4 {\cal E}}$
satisfy the isospin symmetry relations based on the Fierz transformation set
(\ref{Fiers_2E_set}).
Within the nucleon $u$-channel exchange model this set decouples from the
$V_{1 T}$, $A_{1 T}$, $T_{1 T}$ set.
\be
&&
\left. V_{2 {\cal E}}^{VN}(x_1, x_2, x_3, \xi, \Delta^2) \right|_{N(940)} = \Theta_{\rm ERBL}(x_1,x_2,x_3) \frac{1}{(2 \xi)^2} V^p \left(  \frac{x_1}{2 \xi}, \frac{x_2}{2 \xi}, \frac{x_3}{2 \xi}  \right)
K_{2 {\cal E}}^{VN}( \xi, \Delta^2); \nonumber \\ &&
\left. A_{2 {\cal E}}^{VN}(x_1, x_2, x_3, \xi, \Delta^2) \right|_{N(940)} = \Theta_{\rm ERBL}(x_1,x_2,x_3) \frac{1}{(2 \xi)^2} A^p \left(  \frac{x_1}{2 \xi}, \frac{x_2}{2 \xi}, \frac{x_3}{2 \xi}  \right)
K_{2 {\cal E}}^{VN}( \xi, \Delta^2); \nonumber  \\ &&
\left. T_{3 {\cal E}}^{VN}(x_1, x_2, x_3, \xi, \Delta^2)\right|_{N(940)}= -\Theta_{\rm ERBL}(x_1,x_2,x_3) \frac{1}{(2 \xi)^2} T^p \left(  \frac{x_1}{2 \xi}, \frac{x_2}{2 \xi}, \frac{x_3}{2 \xi}  \right)
K_{2 {\cal E}}^{VN}( \xi, \Delta^2); \nonumber  \\ &&
\left. T_{4 {\cal E}}^{VN}(x_1, x_2, x_3, \xi, \Delta^2)\right|_{N(940)}= \Theta_{\rm ERBL}(x_1,x_2,x_3) \frac{1}{(2 \xi)^2} T^p \left(  \frac{x_1}{2 \xi}, \frac{x_2}{2 \xi}, \frac{x_3}{2 \xi}  \right)
K_{2 {\cal E}}^{VN}( \xi, \Delta^2), \nonumber  \\ &&
\ee
where
\be
K_{2 {\cal E}}^{VN}( \xi, \Delta^2)=\frac{f_N}{\Delta^2-M^2} \left( G^V_{VNN}(-2 \xi) + G^T_{VNN}\xi \right).
\ee

\item $V_{2T}$, $A_{2 T}$, $T_{2T}$, $T_{3 T}$ satisfy the isospin symmetry relations based on the Fierz transformation set (\ref{Fiers_2T_set}).
\be
&&
\left. V_{2 T}^{VN}(x_1, x_2, x_3, \xi, \Delta^2) \right|_{N(940)}
 = \Theta_{\rm ERBL}(x_1,x_2,x_3) \frac{1}{(2 \xi)^2} V^p \left(  \frac{x_1}{2 \xi}, \frac{x_2}{2 \xi}, \frac{x_3}{2 \xi}  \right)
K_{2 T}^{VN}( \xi, \Delta^2); \nonumber \\ &&
\left. A_{2 T}^{VN}(x_1, x_2, x_3, \xi, \Delta^2) \right|_{N(940)}
= \Theta_{\rm ERBL}(x_1,x_2,x_3) \frac{1}{(2 \xi)^2} A^p \left(  \frac{x_1}{2 \xi}, \frac{x_2}{2 \xi}, \frac{x_3}{2 \xi}  \right)
K_{2 T}^{VN}( \xi, \Delta^2); \nonumber  \\ &&
\left. T_{2 T}^{VN}(x_1, x_2, x_3, \xi, \Delta^2)\right|_{N(940)}=
 -\Theta_{\rm ERBL}(x_1,x_2,x_3) \frac{1}{(2 \xi)^2} T^p \left(  \frac{x_1}{2 \xi}, \frac{x_2}{2 \xi}, \frac{x_3}{2 \xi}  \right)
K_{2 T}^{VN}( \xi, \Delta^2); \nonumber  \\ &&
\left. T_{3 T}^{VN}(x_1, x_2, x_3, \xi, \Delta^2)\right|_{N(940)}=
-\Theta_{\rm ERBL}(x_1,x_2,x_3) \frac{1}{(2 \xi)^2} T^p \left(  \frac{x_1}{2 \xi}, \frac{x_2}{2 \xi}, \frac{x_3}{2 \xi}  \right)
K_{2 T}^{VN}( \xi, \Delta^2), \nonumber  \\ &&
\ee
where
\be
K_{2 T}( \xi, \Delta^2)= \frac{f_N}{\Delta^2-M^2} \left( G^T_{VNN}   \frac{\xi (1+\xi)}{1-\xi} \right).
\ee

\item $V_{2n}$, $A_{2 n}$, $T_{2n}$, $T_{3 n}$ satisfy the isospin symmetry relations based on the Fierz transformation set (\ref{Fiers_2n_set}).
\be
&&
\left. V_{2 n}^{VN}(x_1, x_2, x_3, \xi, \Delta^2) \right|_{N(940)}
= \Theta_{\rm ERBL}(x_1,x_2,x_3) \frac{1}{(2 \xi)^2} V^p \left(  \frac{x_1}{2 \xi}, \frac{x_2}{2 \xi}, \frac{x_3}{2 \xi}  \right)
K_{2 n}^{VN}( \xi, \Delta^2); \nonumber \\ &&
\left. A_{2 n}^{VN}(x_1, x_2, x_3, \xi, \Delta^2) \right|_{N(940)} = \Theta_{\rm ERBL}(x_1,x_2,x_3) \frac{1}{(2 \xi)^2} A^p \left(  \frac{x_1}{2 \xi}, \frac{x_2}{2 \xi}, \frac{x_3}{2 \xi}  \right)
K_{2 n}^{VN}( \xi, \Delta^2); \nonumber  \\ &&
\left. T_{2 n}^{VN}(x_1, x_2, x_3, \xi, \Delta^2)\right|_{N(940)}= -\Theta_{\rm ERBL}(x_1,x_2,x_3) \frac{1}{(2 \xi)^2} T^p \left(  \frac{x_1}{2 \xi}, \frac{x_2}{2 \xi}, \frac{x_3}{2 \xi}  \right)
K_{2 n}^{VN}( \xi, \Delta^2); \nonumber  \\ &&
\left. T_{3 n}^{VN}(x_1, x_2, x_3, \xi, \Delta^2)\right|_{N(940)}= -\Theta_{\rm ERBL}(x_1,x_2,x_3) \frac{1}{(2 \xi)^2} T^p \left(  \frac{x_1}{2 \xi}, \frac{x_2}{2 \xi}, \frac{x_3}{2 \xi}  \right)
K_{2 n}^{VN}( \xi, \Delta^2), \nonumber  \\ &&
\ee
where
\be
K_{2n}^{VN}(\xi, \Delta^2)= \frac{f_N}{\Delta^2-M^2} \xi   \left(  \frac{1+\xi}{(1-\xi)^2} \frac{  \left(m^2_V-\Delta_T^2\right)}{M^2  }-\frac{1}{1+\xi}\right) G^T_{VNN}.
\ee
\item Finally,
\be
&&
\left. T_{4T}^{VN}(x_1, x_2, x_3, \xi, \Delta^2)\right|_{N(940)}=0;
 \nonumber  \\ &&
\left. T_{4n}^{VN}(x_1, x_2, x_3, \xi, \Delta^2)\right|_{N(940)}=0;
\ee
\ei

The above formulas can be employed both for $I=0$ and $I=1$
vector-meson-to-nucleon TDAs. In the latter case the $u$-channel
nucleon pole exchange contributes only  to the $u$-channel isospin-$\frac{1}{2}$
invariant amplitude
$M^{(\rho N)_{1/2} \, \{1 2 \}}_{\rho \tau \chi}$,
thus populating only  $u$-channel isospin-$\frac{1}{2}$ TDAs.

It is straightforward to check that the $I=0$ and $I=1$
vector-meson-to-nucleon TDAs  computed within the
$u$-channel
nucleon pole exchange model satisfy the set of isospin identities
following from the appropriate isospin symmetry relations
(\ref{Isospin_Id_omegaN})
and
(\ref{Isospin_Id_rhoN}).
The explicit form of these isospin identities  can be established with the
use of the set of the Fierz identities worked out in App.~\ref{App_Isospin}.
For example for the case of $V_{1 {\cal E}}^{VN}$ TDA it reads
\be
&&
V_{1 {\cal E}}^{VN}(x_1,x_2,x_3,\xi, \Delta^2)+ \frac{1}{2}
\left(
V_{1 {\cal E}}^{VN}-A_{1 {\cal E}}^{VN}+T_{1 {\cal E}}^{VN}+T_{2 {\cal E}}^{VN}
\right)(x_3,x_1,x_2,\xi, \Delta^2)
\nonumber \\ &&
+ \frac{1}{2}
\left(
V_{1 {\cal E}}^{VN}-A_{1 {\cal E}}^{VN}+T_{1 {\cal E}}^{VN}+T_{2 {\cal E}}^{VN}
\right)(x_3,x_2,x_1,\xi, \Delta^2)=0.
\ee
The validity of this and all subsequent
isospin identities for $VN$ TDAs within the $u$-channel nucleon exchange model
turns out to be the consequence of the familiar isospin identity for
the leading twist nucleon DAs
\be
2T^p(y_1,y_2,y_3)=(V^p-A^p)(y_1,y_3,y_2)-(V^p-A^p)(y_2,y_3,y_1).
\ee


\begin{thebibliography}{99}

\bibitem{Collins:1996fb}
  J.~C.~Collins, L.~Frankfurt and M.~Strikman,
  Phys.\ Rev.\  D {\bf 56}, 2982 (1997)
  [arXiv:hep-ph/9611433].

\bibitem{Radyushkin:1997ki}
  A.~V.~Radyushkin,
  Phys.\ Rev.\  D {\bf 56}, 5524 (1997)
  [arXiv:hep-ph/9704207].



\bibitem{Frankfurt:1999fp}
  L.~L.~Frankfurt, P.~V.~Pobylitsa, M.~V.~Polyakov and M.~Strikman,
  Phys.\ Rev.\  D {\bf 60} (1999) 014010;
  [arXiv:hep-ph/9901429].

\bibitem{Frankfurt:2002kz}
  L.~Frankfurt, M.~V.~Polyakov, M.~Strikman, D.~Zhalov and M.~Zhalov,
 {\em ``Novel hard semiexclusive processes and color singlet clusters in hadrons''},
  hep-ph/0211263.



\bibitem{Pire:2005ax}
  B.~Pire and L.~Szymanowski,
  Phys.\ Lett.\  B {\bf 622}, 83 (2005)
  [arXiv:hep-ph/0504255].



\bibitem{Lansberg:2007ec}
  J.~P.~Lansberg, B.~Pire and L.~Szymanowski,
  Phys.\ Rev.\ D {\bf 75}, 074004 (2007)
  [Erratum-ibid.\ D {\bf 77}, 019902 (2008)]
  [hep-ph/0701125].

\bibitem{Lansberg:2007se}
  J.~P.~Lansberg, B.~Pire and L.~Szymanowski,
  Phys.\ Rev.\ D {\bf 76}, 111502 (2007)
  [arXiv:0710.1267 [hep-ph]].



\bibitem{Impact1}
   M.~Burkardt,
  Phys.\ Rev.\  D {\bf 62}, 071503, (2000) [hep-ph/0005108].

\bibitem{Impact2}
 J.~P.~Ralston and B.~Pire,
  Phys.\ Rev.\  D {\bf 66}, 111501 (2002)
  [hep-ph/0110075].

\bibitem{Impact3}
  M.~Diehl,
  Eur.\ Phys.\ J.\  C {\bf 25}, 223 (2002)
  [Erratum-ibid.\  C {\bf 31} (2003) 277]  [hep-ph/0205208].


\bibitem{Pire:2013jva}
  B.~Pire, K.~Semenov-Tian-Shansky and L.~Szymanowski,
  Phys.\ Lett.\ B {\bf 724}, 99 (2013)
  [arXiv:1304.6298 [hep-ph]].

\bibitem{Pire:2013tpa}
  B.~Pire, K.~Semenov-Tian-Shansky and L.~Szymanowski,
  Few Body Syst.\  {\bf 55}, 351 (2014)
  [arXiv:1312.7120 [hep-ph]].


\bibitem{Pire:2011xv}
  B.~Pire, K.~Semenov-Tian-Shansky and L.~Szymanowski,
  Phys.\ Rev.\ D {\bf 84}, 074014 (2011)
  [arXiv:1106.1851 [hep-ph]].

\bibitem{Lansberg:2011aa}
  J.~P.~Lansberg, B.~Pire, K.~Semenov-Tian-Shansky and L.~Szymanowski,
  Phys.\ Rev.\ D {\bf 85}, 054021 (2012)
  [arXiv:1112.3570 [hep-ph]].

\bibitem{Lansberg:2012ha}
  J.~P.~Lansberg, B.~Pire, K.~Semenov-Tian-Shansky and L.~Szymanowski,
  Phys.\ Rev.\ D {\bf 86}, 114033 (2012)
  [Erratum-ibid.\ D {\bf 87}, no. 5, 059902 (2013)]
  [arXiv:1210.0126 [hep-ph]].

\bibitem{Kub}
A.~Kubarovskiy,  {\it ``Electroproduction of $\pi^0$ at high momentum transfers in non-resonant region with CLAS''},
JLAB-PHY-12-1415.

\bibitem{Ma:2014pka}
  B.~Ma, B.~Pire, K.~Semenov-Tian-Shansky and L.~Szymanowski,
  EPJ Web Conf.\  {\bf 73}, 05006 (2014)
  [arXiv:1402.0413 [hep-ph]].

\bibitem{Ma2}
 B.~Ma, {\it ``Simulation of electromagnetic channels with ¯PANDA at FAIR''}, PhD thesis, Universit\'{e} de Paris-Sud, 2014.

\bibitem{Singh:2014pfv}
  B.~P.~Singh {\it et al.}  [PANDA Collaboration],
  {\it ``Experimental access to Transition Distribution Amplitudes with the \={P}ANDA experiment at FAIR''},
  arXiv:1409.0865 [hep-ex].



\bibitem{Shanahan:2014tja}
  P.~E.~Shanahan, R.~Horsley, Y.~Nakamura, D.~Pleiter, P.~E.~L.~Rakow, G.~Schierholz, H.~St\"{u}ben and A.~W.~Thomas {\it et al.},
  {\it ``Determination of the strange nucleon form factors,''}
  arXiv:1403.6537 [hep-lat].

\bibitem{Lansberg:2006uh}
  J.~P.~Lansberg, B.~Pire and L.~Szymanowski,
  Nucl.\ Phys.\ A {\bf 782}, 16 (2007)
  [hep-ph/0607130].


\bibitem{Pire:2010if}
  B.~Pire, K.~Semenov-Tian-Shansky and L.~Szymanowski,
  Phys.\ Rev.\ D {\bf 82}, 094030 (2010)
  [arXiv:1008.0721 [hep-ph]].


\bibitem{Chernyak:1984bm}
  V.~L.~Chernyak and I.~R.~Zhitnitsky,
  Nucl.\ Phys.\ B {\bf 246}, 52 (1984).


\bibitem{Kroll:1995pv}
  P.~Kroll, M.~Schurmann, P.~A.~M.~Guichon,
  Nucl.\ Phys.\  {\bf A598}, 435-461 (1996).
  [hep-ph/9507298].


\bibitem{HuberPrivate}
G. Huber, {\it private communication}.


\bibitem{Stef}
N.~G.~Stefanis, Eur. Phys. J. direct C {\bf 7}, 1 (1999).

\bibitem{Lepage:1980fj}
  G.~P.~Lepage and S.~J.~Brodsky,
  Phys.\ Rev.\ D {\bf 22}, 2157 (1980).
  
  
\bibitem{Bolz:1996sw}
  J.~Bolz and P.~Kroll,
  Z.\ Phys.\ A {\bf 356}, 327 (1996)
  [hep-ph/9603289].

\bibitem{Braun:2006hz}
  V.~M.~Braun, A.~Lenz and M.~Wittmann,
  Phys.\ Rev.\ D {\bf 73}, 094019 (2006)
  [hep-ph/0604050]. 
  
\bibitem{Lenz:2009ar}
  A.~Lenz, M.~Gockeler, T.~Kaltenbrunner and N.~Warkentin,
  Phys.\ Rev.\ D {\bf 79}, 093007 (2009)
  [arXiv:0903.1723 [hep-ph]].   
  
\bibitem{Braun:2014wpa}
  V.~M.~Braun, S.~Collins, B.~Gl\"{a}ssle, M.~G\"{o}ckeler, A.~Sch\"{a}fer, R.~W.~Schiel, W.~S\"{o}ldner and A.~Sternbeck {\it et al.},
  Phys.\ Rev.\ D {\bf 89}, no. 9, 094511 (2014)
  [arXiv:1403.4189 [hep-lat]].  
  

\bibitem{King:1986wi}
  I.~D.~King and C.~T.~Sachrajda,
  Nucl.\ Phys.\  B {\bf 279}, 785 (1987).
  
\bibitem{Chernyak:1987nv}
  V.~L.~Chernyak, A.~A.~Ogloblin and I.~R.~Zhitnitsky,
  Z.\ Phys.\  C {\bf 42}, 583 (1989)
  [Yad.\ Fiz.\  {\bf 48}, 1398 (1988)]
  [Sov.\ J.\ Nucl.\ Phys.\  {\bf 48}, 889 (1988)].
  

\bibitem{Gari:1986ue}
  M.~Gari and N.~G.~Stefanis,
  Phys.\ Lett.\  B {\bf 175}, 462 (1986).












\bibitem{Radyushkin:1990te}
  A.~V.~Radyushkin,
  Nucl.\ Phys.\ A {\bf 532}, 141 (1991).
  
\bibitem{Anikin:2013aka}
  I.~V.~Anikin, V.~M.~Braun and N.~Offen,
  Phys.\ Rev.\ D {\bf 88}, 114021 (2013)
  [arXiv:1310.1375 [hep-ph]].


\bibitem{Grein:1979nw}
  W.~Grein and P.~Kroll,
  Nucl.\ Phys.\ A {\bf 338}, 332 (1980).


\bibitem{Dumbrajs:1983jd}
  O.~Dumbrajs, R.~Koch, H.~Pilkuhn, G.~c.~Oades, H.~Behrens, J.~j.~De Swart and P.~Kroll,
  Nucl.\ Phys.\ B {\bf 216}, 277 (1983).

\bibitem{EMPworkshop}
G.~Huber, {\em ``Backward angle production of $\omega$ and $\phi$ mesons in Hall C''}, a talk presented at
Exclusive Meson Production Workshop, January 22-24 2015, JLab.


\bibitem{Mergell:1995bf}
  P.~Mergell, U.~G.~Meissner and D.~Drechsel,
  Nucl.\ Phys.\ A {\bf 596}, 367 (1996)
  [hep-ph/9506375].

\bibitem{Meissner:1997qt}
  U.~G.~Meissner, V.~Mull, J.~Speth and J.~W.~van Orden,
  Phys.\ Lett.\ B {\bf 408}, 381 (1997)
  [hep-ph/9701296].

\bibitem{Hohler:1976ax}
  G.~Hohler, E.~Pietarinen, I.~Sabba Stefanescu, F.~Borkowski, G.~G.~Simon, V.~H.~Walther and R.~D.~Wendling,
  Nucl.\ Phys.\ B {\bf 114}, 505 (1976).


\bibitem{Pire:2004ie}
  B.~Pire and L.~Szymanowski,
  Phys.\ Rev.\ D {\bf 71}, 111501 (2005)
  [hep-ph/0411387].



\end{thebibliography}
\end{document}